\documentclass[10pt]{amsart}
\usepackage{amsmath,amsfonts,amssymb,amsthm,amscd,latexsym,euscript,xypic, verbatim}
\usepackage{amstext}
\usepackage{graphicx}
\usepackage[active]{srcltx}
\usepackage[all]{xy}
\usepackage{graphics}
\usepackage{color}
\usepackage[utf8]{inputenc}
\usepackage{mathtools}
\usepackage{tikz}
\usepackage{dsfont}

\swapnumbers
\theoremstyle{plain}

\newtheorem*{thm1.2}{(1.2) Theorem}
\newtheorem*{thm1.3}{(1.3) Theorem}
\newtheorem*{thm1.4}{(1.4) Theorem}
\newtheorem*{propA*}{Proposition A}
\newtheorem*{propB*}{Proposition B}
\newtheorem*{thmC*}{Theorem C}
\newtheorem*{propD*}{Proposition D}
\newtheorem{prop}{Proposition}[section]

\theoremstyle{definition}
\newtheorem{point}[prop]{}

\newtheorem{Def}[prop]{Definition}

\newtheorem*{Def*}{Definition}

\newtheorem*{notation*}{Notation}
\newtheorem*{question*}{Question}

\newcommand{\D}{\textbf D}
\renewcommand{\P}{\textbf P}

\newcommand{\N}{\mathbf N}

\newcommand{\R}{\mathbf R}

\newcommand{\ra}{\rightarrow}

\xyoption{2cell}

\makeatletter
\newcases{mycases}{\quad}{%
	\hfil$\m@th\displaystyle{##}$}{$\m@th\displaystyle{##}$\hfil}{\lbrace}{.}
\makeatother

\title{Generalized persistence analysis based on stable rank invariant}

\author{Henri Riihim\"{a}ki*}
\thanks{*Tampere University of Technology}
\author{Wojciech Chacholski**}
\thanks{**KTH Royal Institute of Technology}
\date{\today}

\begin{document}
\maketitle

\begin{abstract}
We believe three ingredients are needed for  further progress in persistence and its use: 
invariants not relying on decomposition theorems to go beyond 1-dimension, outcomes suitable for statistical analysis and a setup adopted for supervised and machine learning. Stable rank, a continuous invariant for multidimensional persistence, was introduced in \cite{Chacholski2015}. 
In the current paper we continue this work by demonstrating how one builds an efficient computational pipeline around this invariant and uses it in inference in case of one parameter. We demonstrate some computational evidence of the statistical stability of stable rank. We also show how our framework can be used in supervised learning.
\end{abstract}

\section{intro}
Topological data analysis (TDA) and particularly its subfield persistent homology aims at quantifying topological features in data sets \cite{CarlssonPointCloud,Oudot,Roadmap}. Finding connected components of data is the aim of clustering. Loop structure might signal about a recurrent dynamics of the phenomenon that produced the data. Various dimensional voids can  mark for example lack of information and connectivity or insufficient data collection. Finding such voids in data sets has aroused interest in different areas of data analysis community, see for example \cite{BigHoles} and references therein. As noted in \cite{BigHoles}, voids can also indicate non-allowed combinations of feature values.  

The fundamental problem of using persistent homology in data analysis and machine learning is the difficulty of performing statistical analyses with its outcome. This is due to the fact that it simply makes no sense to take averages of homologies of some collection of spaces. Different functional reformulations of persistence output have been proposed to overcome this difficulty \cite{Bubenik2015,PersImages,PersTerrace}. These, however, are based on the standard barcode decomposition or persistence diagram and thus are only limited to 1-dimensional setting.

Our aim for this paper is to illustrate an entirely new approach of extracting homological information that is suitable for validation and statistical analysis, effective in supervised learning as well. We emphasize that the approach in general does not rely on any algebraic decomposition of persistence and is thus applicable to multiparameter persistence. In this article we only briefly explain the mathematical fundamentals of our method, referring the reader to \cite{Chacholski2015,StableInvariants} for more in depth mathematical discussions. The  focus is on demonstrating how to use our computational pipeline to extract and organize homological information and explore inference in explicit data sets. Due to computational reasons we work on 1-dimensional setting.

Persistent homology works by incrementally building and analyzing geometrical structure on data. Traditional view in persistence has been that topological features with long lifetimes are of importance and small scale features are to be considered as noise. This view, however, is challenged by many recent studies showing that smaller features carry important information: study of brain artery trees in \cite{Bendich2016}, functional networks of \cite{Stolz2016}, relation of fullerene curvature energy with persistence in \cite{XiaWei_CryoEM} and analysis of protein structure in \cite{XiaWei_protein}. See also \cite{Hiraoka_amorphous} where small loops in atomic configurations of amorphous silica glass were found to provide explanation to observed diffraction peak. Further examples that we also study in this paper are point processes on a unit square. Here no large scale structure is to be expected and differences are found in the small scale clustering and looping of points. Consider also the following example. Let our data set consist of a group of people. We impose on  this set  a $t$-parameterized relation $S_t$ where $x S_t y$ if persons $x$ and $y$ have known each other at most $t$ years for $t \in [0,\infty)$. Note that $S_t$ is not a metric. 
We can use $t$ as the persistence parameter. For the Vietoris-Rips construction at scale $t$, $k+1$ persons form a $k$-simplex
if all the persons have known each other pairwise at most $t$ years. Now at some value of $t_1$ four people might form a 3-simplex $\sigma$ denoting a subgroup or a cluster of people in our social relation $S_t$. At some later value $t_2$ this cluster might connect to some other cluster ending the independent existence of $\sigma$. Even though the lifetime $t_2 - t_1$ of $\sigma$ could be small, it might still indicate important information in the social relations of our data.

It thus depends on the analysis at hand what is to be taken as topological noise. The concept of noise was formalized in \cite{Chacholski2015} for general multiparameter persistence, leading to the definition of our invariant, \emph{stable rank}. In one-dimensional persistence, the stable rank is a (non-increasing) function counting the number of features, or bars of the barcode decomposition, at different scales. Our pipeline is further based on so called \emph{persistence contours}. With flexibility of using different persistence contours, we are able to produce a whole range of stable ranks capturing different aspects of data topology. By casting the output from persistence pipeline into our functional invariant one is able to further connect persistence with statistics and machine learning. Persistence contours appeared independently also in \cite{BubenikDeSilva2015} under the name superlinear families but with different purpose of defining abstract interleavings between generalized persistence modules. We will elaborate on this in Section~\ref{standard_persistence}. We detail our \emph{generalized persistence} pipeline in Sections~\ref{generalized_persistence and persistence contours} and~\ref{section Stable rank}. Section \ref{stability} gives computational evidence of the statistical stability of our approach and how taking means of our invariant can be used to stabilize persistence output. In section \ref{contour_usage} we show with concrete case studies how this pipeline leads to a more flexible analysis and feature extraction of persistence in supervised classification. To generate relevant  barcodes needed for our calculations we used the Ripser software \cite{Ripser}.

\section{Standard and generalized persistence}\label{standard_persistence}
A typical input for persistence analysis is a finite metric space $(\mathcal{M},d)$. The first step of the analysis is to encode the metric information into a simplicial complex parametrized by
the poset of non-negative reals (referred to as  scales)  denoted by $\R$. In this paper  the Vietoris-Rips construction
is used  for that purpose, which  at scale $t\in \R$  is a simplicial complex $\text{VR}_t(\mathcal{M})$ whose $k$-simplices are subsets of $\mathcal{M}$ consisting of $k+1$ points which are pairwise at most distance $t$ from each other. For example, two data points $p_1 ,p_2$ in $\mathcal{M}$ connect to an edge, or a 1-simplex, when $d(p_1,p_2) \le t$. Increasing $t$ gives a \emph{filtration} so that for $\cdots\le a \le b \le \cdots $ in $\R$:
\begin{equation*}
\cdots \subseteq \text{VR}_a(\mathcal{M}) \subseteq \text{VR}_b(\mathcal{M}) \subseteq \cdots.
\end{equation*} 
Note $\text{VR}_{\bullet}(\mathcal{M})$ does not add or forget any information about $(\mathcal{M},d)$ and hence is as complex as  the metric space itself.
Simplification is therefore necessary and this is the purpose of the second step of persistence analysis. In this step 
the $n$-th homology (with coefficients in a chosen field $K$, for example ${\mathbf F}_2$) is applied to the simplicial complexes  in the Vietoris-Rips filtration producing an $\R$-parametrized vector space (here the arrows denote linear maps):
\begin{equation*}
\cdots \ra H_n (\text{VR}_a(\mathcal{M}),K) \ra H_n (\text{VR}_b(\mathcal{M}),K) \ra \cdots.
\end{equation*}
The obtained result is not an arbitrary $\R$-parametrized vector space. Its values $H_{n,t}:=H_n (\text{VR}_t(\mathcal{M}),K)$, at different scales, are finite dimensional and there are finitely many numbers $0<t_0<\cdots <t_k$ in $\R$  such that the map $H_{n,a}\to H_{n,b}$  may fail  to be an isomorphism only if  $a<t_i\leq b$ for some $i$. Such parametrized vector spaces are called tame \cite{Chacholski2015}. In this parameterized sequence the dimensions of homology vector spaces encode topological information: $H_{0,t}$ effectively measuring the number of connected components, $H_{1,t}$ measuring the number one-dimensional holes and $H_{k,t}$ those of $k$-dimensional voids at scale $t$.

An essential fact needed for standard persistence analysis is the decomposition theorem which states that any tame $\R$-parametrized vector space is a direct sum of indecomposables called bars and the collection of bars in such a decomposition is unique \cite{Carlsson2005}. Bars are enumerated by pairs of numbers $b<d$ in $\R$. The bar $[b,d)$ at scale $t$ is either the  one dimensional vector space $K$, if $b\leq t<d$, or the zero vector space otherwise. The maps between any non-zero values  in a bar are isomorphisms.
Typically the bar $[b,d)$ is   illustrated as follows:
$$\cdots 0 \overset{0}{\longrightarrow} K_b \overset{\mathds{1}}{\longrightarrow} K_d \overset{0}{\longrightarrow} 0  \cdots.$$

In the standard approach the actual data analysis step involves assigning importance to the bars in a decomposition of $H_{n,\bullet}$. That is why the decomposition theorem has been so fundamental in persistence. Our approach is essentially different since to define our invariants we do not use the decomposition theorem. We use this theorem for effective calculations but not for conceptual reasons. 

How to extract and organize  relevant information about  $H_{n,\bullet}$? We claim that the answer should depend on the data analysis task at hand. Connecting persistence analysis to machine learning is important and one of our aims in this article is to illustrate how our approach can be used to improve classification accuracy in the setting of supervised learning by appropriately selecting features from persistence output. As sets of intervals, statistics on bars is difficult. Using our invariant we are able to study averages and perform statistical analysis.

Our pipeline is based on  flexibility of choosing and adopting different metrics on tame parametrized vector spaces. Any such choice leads to a different stable invariant and we explore this freedom by selecting a more suitable one relevant to a given task.
The most fundamental invariant of a tame parametrized vector space  $F_{\bullet}$ is its minimal number of generators  called also the {\em rank} and denoted by  $\text{rank}(F_{\bullet})$. It corresponds to the number of bars in the bar decomposition of $F_{\bullet}$. For a choice of a pseudo-metric $\mu$ (see~\cite{StableInvariants}) on tame parametrized vector spaces, define the {\em stable rank} to be the function $\widehat{\text{rank}}(F_{\bullet})\colon \R\to \R$  such that:
\[\widehat{\text{rank}}(F_{\bullet})(t):=\text{min}\{\text{rank}(G_{\bullet})\ |\ \mu(G_{\bullet},F_{\bullet})\leq t\}\]
Thus $\widehat{\text{rank}}(F_{\bullet})(t)$ is the smallest rank  among the ranks of tame parametrized vector spaces in the disc of radius $t$ around $F_{\bullet}$. Note that the decomposition theorem plays no role in the definition of the stable rank. The essential  observation in \cite{Chacholski2015} is that assignment $F_{\bullet}\mapsto \widehat{\text{rank}}(F_{\bullet})$ is a continuous operation between tame parametrized vector spaces and measurable functions from $\R$ to $\R$.

A choice of a pseudo-metric $\mu$ on tame parametrized vector spaces leads therefore to the following stable signature of a finite metric space $(\mathcal{M},d)$:
$$\widehat{\text{rank}}_n(\mathcal{M}):=\widehat{\text{rank}}(H_n (\text{VR}_\bullet(\mathcal{M}),K) )\colon\R\to\R$$
This signature summarizes the topological features in the Vietoris-Rips construction captured by $n$-th homology. Usefulness of this signature depends on our ability of choosing many pseudo-metrics for which we are able to calculate corresponding stable ranks. Noise systems in \cite{Chacholski2015} were introduced exactly for the purpose of defining pseudo-metrics on tame parametrized vector spaces. For implementing on a computer so called simple noise systems (\cite{StableInvariants}, \cite{Chacholski2015}) are much more convenient. The reason is that simple noise systems are parametrized by \emph{persistent contours}. Since the intention in this paper is to illustrate the usefulness of our approach, instead of explaining the theory behind noise systems, we focus on discussing only contours and how they directly can be used to calculate the different stable ranks. This is the content of the next section. We believe that it is important however to be aware of the relation between contours and noise systems.

Persistence contours appeared independently already in \cite{BubenikDeSilva2015} under the name \emph{superlinear families}. A superlinear family is an indexed set of translations of some preordered set $\P$, $\textbf{Trans}_\P$, or a function $\Omega : \R \ra \textbf{Trans}_\P,$ satisfying $\Omega_{\epsilon_2} \circ \Omega_{\epsilon_1} \le \Omega_{\epsilon_1 + \epsilon_2}$, whenever $\epsilon_1, \epsilon_2 \ge 0$. In \cite{BubenikDeSilva2015} persistence theory was expanded to diagrams indexed by $\P$ and constructed by other tools than homology, thus taking values in any appropriate category $\D$. Resulting persistence modules were called \emph{generalized persistence modules}. Superlinear families were then used to define interleaving distance between generalized persistence modules. Since then generalized persistence modules have gathered some interest, see \cite{Puuska} and \cite{MeehanMeyer}.

We emphasize our different point of view. Superlinear families enabled abstract study of interleaving metrics between generalized persistence modules. In a sense this corresponds to pre-composing, or re-parameterizing, the persistence functor $G: \P \ra \D$ with $\Omega_\epsilon$. Pre-composing with a re-parameterization imposes a priori choice of scale of features. First computing persistence and then applying persistence contour allows us to study the effect of different contours in the analysis at hand and produce range of invariants emphasizing different aspects of data, e.g. for a classification task. We call our approach \emph{generalized persistence}, signifying the generalization of persistence analysis pipeline, in contrast to generalized persistence modules, there the emphasis being in the abstract study of persistence functors themselves. 

\section{Persistence contours}\label{generalized_persistence and persistence contours}
\begin{Def} \label{perscont}
	A {\em persistence contour} is a function $C: \R \times \R \ra \R$  satisfying  the following conditions  for all $v, w \in \R$ and $\epsilon, \tau \in \R$:
	\begin{enumerate}
		\item if $v \leq w$ and $\epsilon \leq \tau$, then $C(v,\epsilon) \leq C(w,\tau)$;
		\item $v \leq C(v,\epsilon)$ and $C(C(v,\epsilon),\tau) \leq C(v,\epsilon + \tau)$.
	\end{enumerate}
\end{Def}

The first condition of~\ref{perscont} makes sure that a contour  preserves the poset structures.  The second  one can be depicted graphically as:
\[\xymatrix{
	\bullet\ar@{|->}@/^15pt/[rr]^-{C(\bullet,\epsilon)}\ar@{|->}@/_25pt/[rrrrrr]_-{C(\bullet,\epsilon+\tau)} &\leq & \bullet \ar@{|->}@/^15pt/[rr]^-{C(\bullet,\tau)}& \leq & \bullet &\leq & \bullet
}\]
Note that  if  the  inequalities in  the second condition are replaced by  equalities,  then 
the   contour  describes an action  of the  additive monoid  $(\R,+)$ on the poset $\R$.

For contours to be useful as tools in data analysis we need methods to produce them. We now present several of them  along with examples. This greatly enlarges \cite{BubenikDeSilva2015}, in which the authors only give the standard translation $v+\epsilon$ as a concrete example of superlinear families. 

Definition \ref{perscont} gives  three functional inequalities implicitly characterizing   persistence contours. The last  inequality however makes it difficult to give explicit formulas for  persistence contours. We can however make initial guesses for the  form of a contour and then try to find a formula satisfying the requirements  of Definition~\ref{perscont}.

\begin{point}{\bf Exponential contour.}
	Consider  a function of the form  $C(v,\epsilon) = f(\epsilon)v.$ 
	If $f:\R \ra \R$ is  non-decreasing and  at least $1$ at zero, then $C$ satisfies the first two inequalities   of~\ref{perscont}. The third inequality  is equivalent to:
	\[C(C(v,\epsilon),\tau) = C(f(\epsilon)v,\tau) = f(\tau)f(\epsilon)v \leq f(\epsilon + \tau)v = C(v,\epsilon + \tau)\]
	For instance, since $e^\tau e^\epsilon = e^{\epsilon + \tau}$, the function
	$C(v,\epsilon) =e^\epsilon v$ is a persistance contour. In fact we could choose any positive base number $r$ other than $e$.  Such contours are called {\em exponential}. 
\end{point}

\begin{point}{\bf Standard contour.}\label{standardcont}
	Consider  a function of  the form $C(v,\epsilon) = v + f(\epsilon).$  If $f$ is non-decreasing, then $C$ satisfies  the first two inequalities of~\ref{perscont}. The third inequality gives  $v + f(\epsilon) + f(\tau) \leq v + f(\epsilon + \tau)$. Thus for $C$ to be a contour, $f\colon \R\to\R$ should  be  superlinear  i.e.\  a function satisfying  $f(\epsilon) + f(\tau) \leq f(\epsilon + \tau)$. For  example  $C(v,\epsilon) = v + \epsilon$ is a contour. It is called the  {\em standard contour}. Another  example is the {\em parabolic contour} $C(v,\epsilon) = v + \epsilon^2$. 
\end{point}

Contours can also be described by certain integral equations. Let $f: \R \ra \R$ be a measurable function with strictly positive values referred to as {\em density}. 

\begin{point}{\bf Distance type.}
	Since $f$ has strictly positive values, for any $v$ and $\epsilon$ in $\R$, there is a unique number
	$C(v,\epsilon)\geq v$  for which: 
	$$\epsilon=\int_{v}^{C(v,\epsilon)}f(x) dx.$$
	Additivity of integrals gives:
	\[\epsilon+\tau=\int_{v}^{C(v,\epsilon)}f(x) dx+\int_{C(v,\epsilon)}^{C(C(v,\epsilon),\tau)}f(x) dx=
	\int_{v}^{C(C(v,\epsilon),\tau)}f(x) dx\]
	which  implies  $C(C(v,\epsilon),\tau)=C(v,\epsilon+\tau)$. 
	The inequality $C(v,\epsilon)\leq C(w,\tau)$,  for $v\leq w$ and $\epsilon\leq \tau$, is a consequence of
	monotonicity of integrals.  The function $C$ is therefore a contour. It is called of {\em  distance type} as it describes 
	the distance  needed  to move from $v$ to the right  in order for the area under the graph of $f$ to reach $\epsilon$.
	
	If   density   is  the  constant  function $1$, then $\epsilon = \int_{v}^{C(v,\epsilon)}dx = C(v,\epsilon) - v$ and thus $C(v,\epsilon) = v+ \epsilon$ is the standard contour (see \ref{standardcont}).
\end{point}

\begin{point}{\bf Shift type.}\label{shift type}
	Another way of arriving at a persistence contour is by integrating density function as follows.
	For $v$ in $\R$, choose the unique $y$ such that $v=\int_{0}^{y}f(x)dx$ and define: 
	$$C(v,\epsilon):=\int_{0}^{y+\epsilon}f(x)dx$$
	Monotonicity of integrals implies that $C$ satisfies the  first two inequalities of~\ref{perscont}.
	Since $v=\int_{0}^{y}f(x)dx$  and $C(v,\epsilon)=\int_{0}^{y+\epsilon}f(x)dx$, by definition:
	\[C(C(v,\epsilon),\tau)=\int_{0}^{y+\epsilon+\tau}f(x)dx=C(v,\epsilon+\tau)\]
	The function $C$ is therefore a contour. By writing $C(v,\epsilon)=v+\int_{y}^{y+\epsilon}f(x)dx, \ v=\int_{0}^{y}f(x)dx,$ we see that $C$ is a translation of $v$ by the $\epsilon$-step integral of the density. Therefore it is called of {\em  shift type}.
	
	If the density is the constant function $1$, then $v=\int_{0}^{v}dx$ and hence $C(v,\epsilon)=\int_{0}^{v+\epsilon}dx=v+\epsilon$ is the standard contour.
\end{point}

\section{Stable rank}\label{section Stable rank}
In this section we recall how to  describe  the stable rank $\widehat{\text{rank}}(F_\bullet)\colon \R\to \R$   directly from a contour without passing through the associated metric and minimizing over the induced discs around $F_\bullet$
(see Section \ref{standard_persistence}).

\begin{Def}\label{stablerankfunction}
	Let  $\oplus_{i\in I} [b_i,d_i)$ be the  bar 
	decomposition of a tame $\R$-parametrized vector space $F_{\bullet}$.  
	Then {\em stable rank} function $\widehat{\text{rank}}(F_\bullet)\colon \R\to \R$  is  defined as follows:
	\[\widehat{\text{\rm rank}}(F_\bullet)(\epsilon)=|\{[b_i,d_i)\ |\ C(b_i,\epsilon)<d_i\}|\]
\end{Def}
The values of $\widehat{\text{rank}}(F_\bullet)\colon \R\to \R$ 
count  certain bars in the bar decomposition of $F_\bullet$ and as such  are natural numbers.
The function $\widehat{\text{rank}}(F_\bullet)\colon \R\to \R$ is therefore simple and non-increasing (by $C$ being non-decreasing, see the first condition of \ref{perscont}) and hence measurable. Averages and limits of such functions are again non-increasing and measurable, however no longer with values in natural numbers.

\begin{point}{\bf Metrics for stable ranks.}
	A fundamental fact  that makes the stable rank useful for data analysis is its continuity. In \cite{StableInvariants} it was proved that the operation $F_{\bullet}\mapsto (\widehat{\text{rank}}(F_\bullet)\colon \R\to \R)$ is  continuous  with respect to the metric induced by the contour on the tame $\R$-parametrised vector spaces 
	and either the integral or interleaving distance on measurable functions from $\R$ to $\R$. Recall that the {\em integral} and {\em interleaving distances} between two measurable functions $p,q\colon \R\to \R$ are defined, respectively, as:
	\begin{align*}
	d_{\int}(p,q)&:=\int_0^\infty|p(x)-q(x)|dx \\
	d_{\bowtie} (p,q)&:= \text{inf}\{\epsilon \, | \, p(x) \ge q(x+\epsilon) \ \text{and} \ q(x) \ge p(x+\epsilon) \ \text{for all} \ x \in \R\}
	\end{align*}
	Although we refer to \cite{StableInvariants} for the proof of  the continuity of the stable rank, we will illustrate this  stability behaviour on explicit examples in the following sections.
\end{point}

\begin{point}{\bf Visualizing topological features and contours.}\label{contour_visualization}
	A standard way to visualize the bar decomposition of a tame $\R$-parametrized vector space is by so called persistence diagram in a $(\text{birth}, \text{death})$-coordinate system. The bar decomposition can be equally characterized by the birth and length values of its bars. We can transform persistence diagram to this presentation, which we call {\em persistence stem plot}, by simple change of coordinates
	$(b,d) \mapsto (b,d-b),$ for a $(\text{birth}, \text{death})$-pair $(b,d)$  in a persistence diagram. We find it helpful to visualize pairs $(b,d-b)$ in a $(\text{birth},\text{length})$-plot as vertical stems. Taking into account multiplicity of more than one bar having the same birth value we extend the domain of the stem plot to $\R \times \N$, where $\N$ is used to number bars with the same birth.
	
	For a fixed $\epsilon$, the relation $C(b,\epsilon) < d$ in the Definition \ref{stablerankfunction} of the stable rank describes an area above the curve $\gamma_{\epsilon}(b)=(b,C(b,\epsilon))$ in the $(\text{birth}, \text{death})$-plane. Setting $C(b,\epsilon) = d$ and applying the transformation above, we get a curve $\widehat{\gamma}_{\epsilon}(b)=(b,C(b,\epsilon)-b)$. Since such curves are typically called contour lines,  hence the name persistence contours in \ref{perscont}.
	
	The left plot in Figure \ref{fig_contours_use} illustrates a persistence stem plot along with contour lines for $\epsilon=0,1,2,3$ of the standard and parabolic contours.
	Stem plot and contour lines make it easy to understand how the persistence contour selects features for the associated stable rank. The right plot in Figure \ref{fig_contours_use} depicts these stable ranks. The stable rank for the standard contour (the red graph on the right) gives very similar information as the stem plot, counting the bars in different lengths and the two high persistence features are clearly shown. The stable rank for the parabolic contour (the blue graph on the right) on the other hand discards the two features, their persistence not being large enough to be picked up by the invariant, relatively magnifying the smaller persistence features. If the features would have been large enough the parabolic contour would have indicated this through the stable rank having larger support. Integral distance between these stable ranks is 13 while the interleaving distance is 6.
	\begin{figure}[!h]
		\centering
		\includegraphics[scale=0.65,clip]{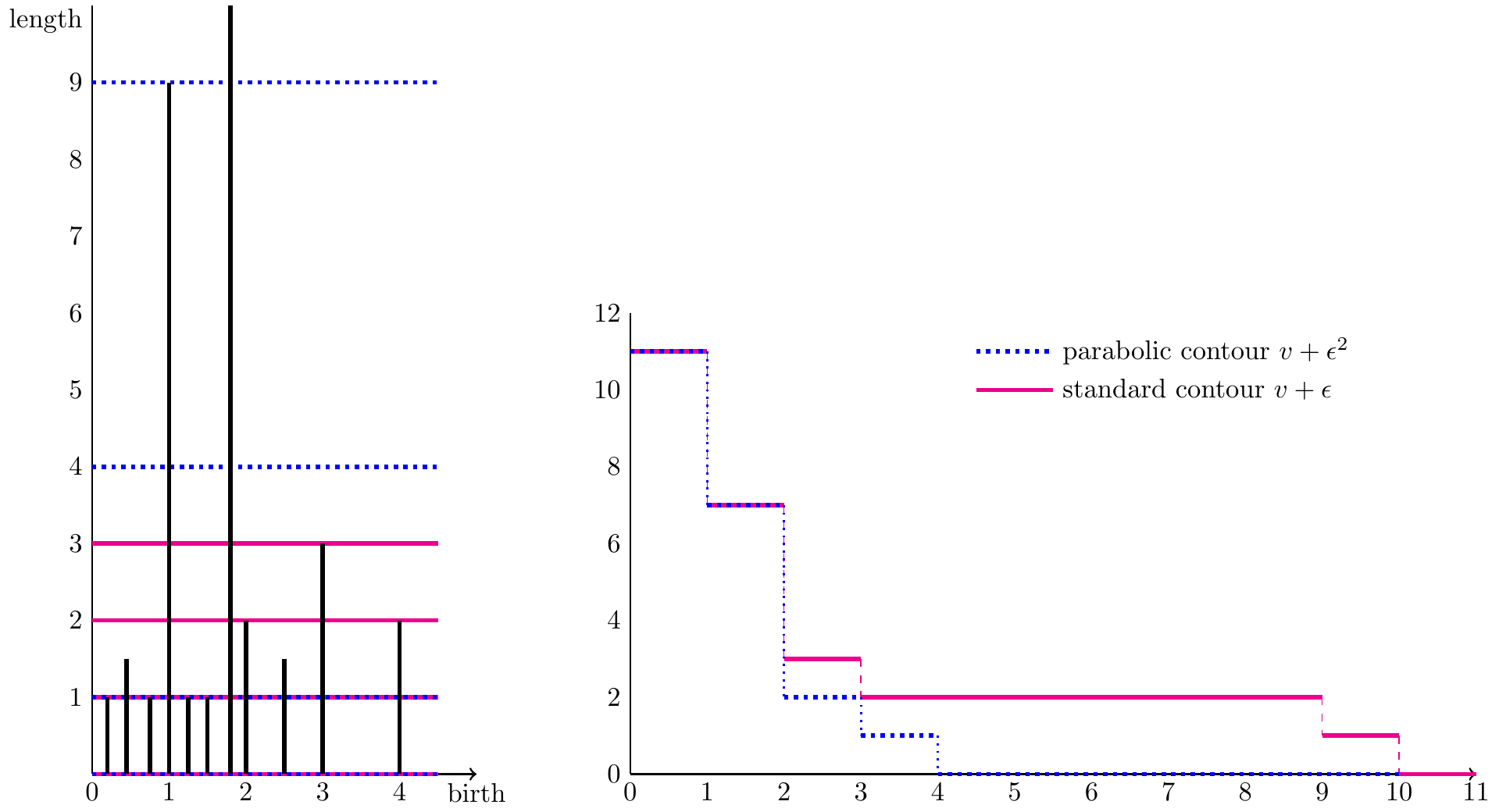}
		\caption{}
		\label{fig_contours_use}
	\end{figure}
\end{point}

\section{Stability}\label{stability}
In this section we consider only the standard contour (see~\ref{standardcont}).
For this contour, our aim  is to   illustrate stability and limit properties of the associated stable ranks:
\[\widehat{\text{rank}}_0(-):=\widehat{\text{rank}}(H_0 (\text{VR}_\bullet(-),{\mathbf F}_2) )\ \ \ \ \ \ \ 
\widehat{\text{rank}}_1(-):=\widehat{\text{rank}}(H_1 (\text{VR}_\bullet(-),{\mathbf F}_2) )\]

\begin{figure}
	\begin{center}
		\begin{minipage}{0.99\textwidth}
			\includegraphics[width=0.142\textwidth]{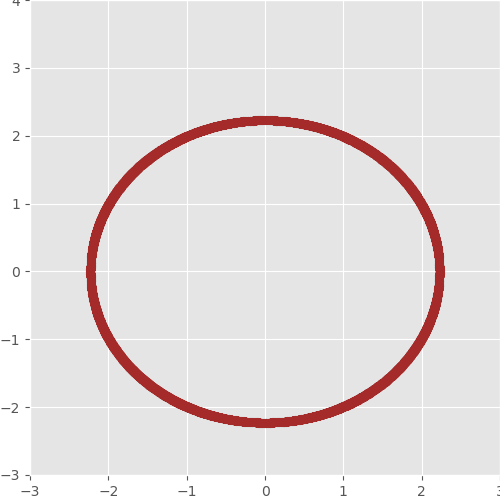}
			\hfill
			\includegraphics[width=0.142\textwidth]{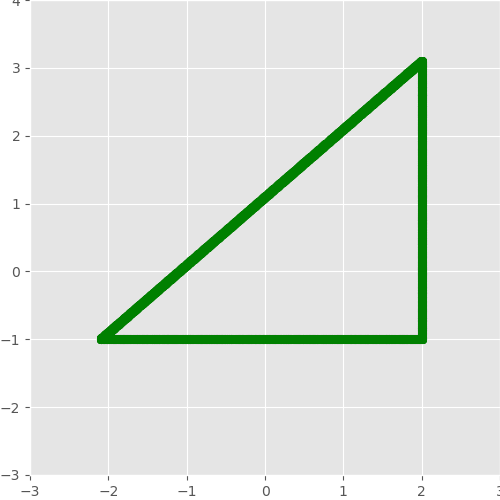}
			\hfill
			\includegraphics[width=0.142\textwidth]{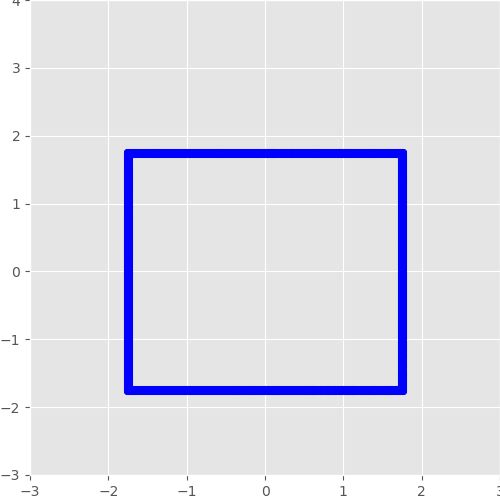}
			\hfill
			\includegraphics[width=0.142\textwidth]{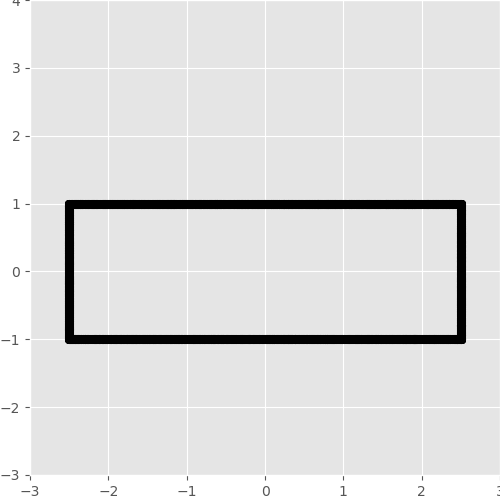}
			\hfill
			\includegraphics[width=0.142\textwidth]{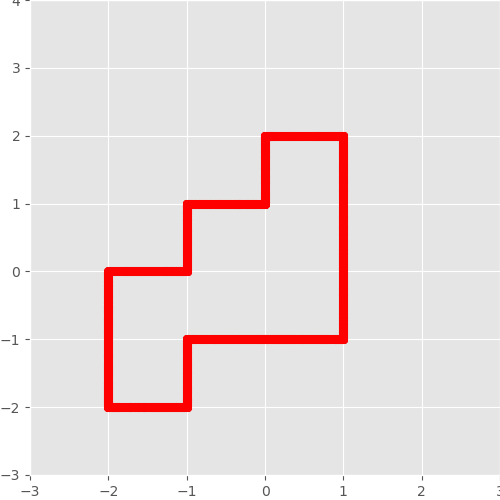}
			\hfill
			\includegraphics[width=0.142\textwidth]{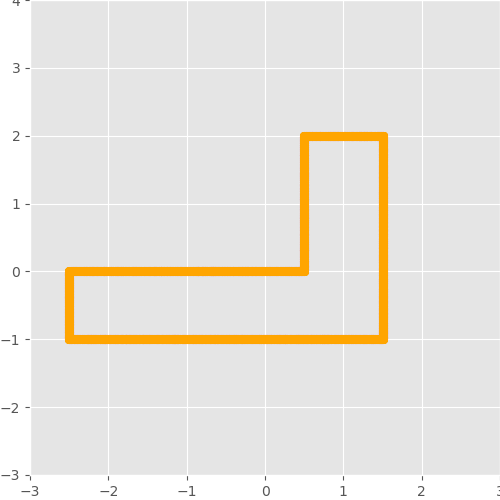}
		\end{minipage}
	\end{center}
	\caption{Closed curves in  Euclidean plane. All curves are isometric to a circle with circumference 14.}
	\label{fig_plane_curves}
\end{figure}

Consider the closed curves in the Euclidean plane shown in Figure \ref{fig_plane_curves}. All these curves are  isometric to a circle with length  $14$. 

\begin{figure}
	\begin{center}
		\begin{minipage}{0.99\textwidth}
			\includegraphics[width=0.142\textwidth]{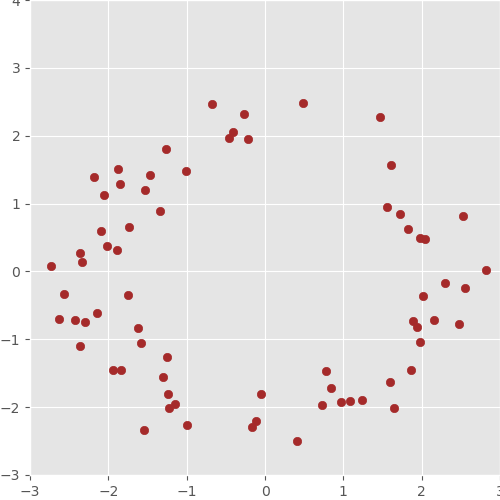}
			\hfill
			\includegraphics[width=0.142\textwidth]{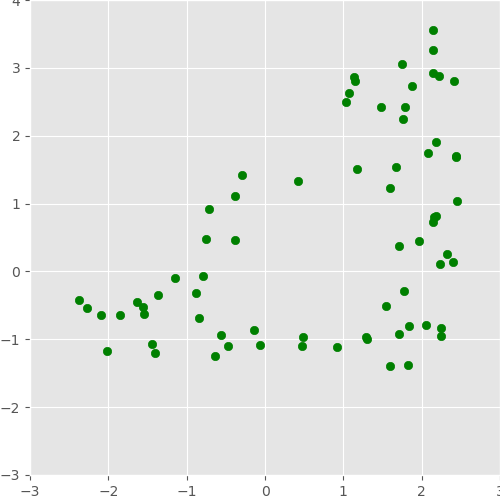}
			\hfill
			\includegraphics[width=0.142\textwidth]{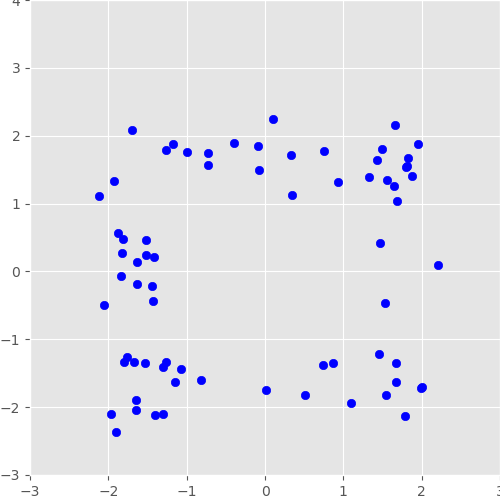}
			\hfill
			\includegraphics[width=0.142\textwidth]{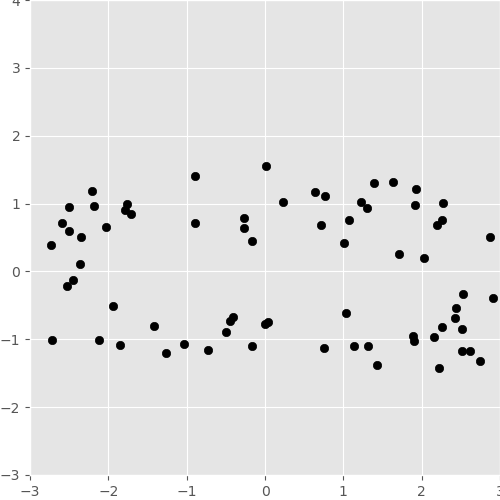}
			\hfill
			\includegraphics[width=0.142\textwidth]{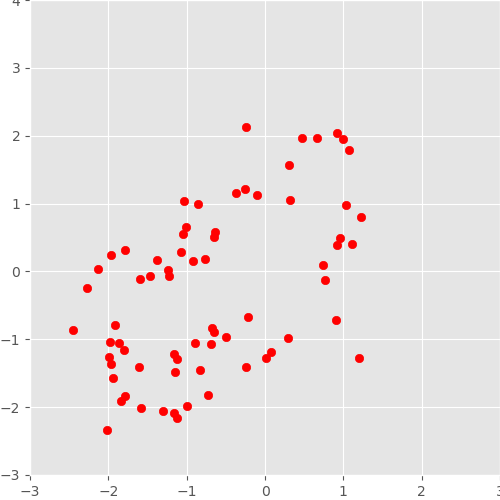}
			\hfill
			\includegraphics[width=0.142\textwidth]{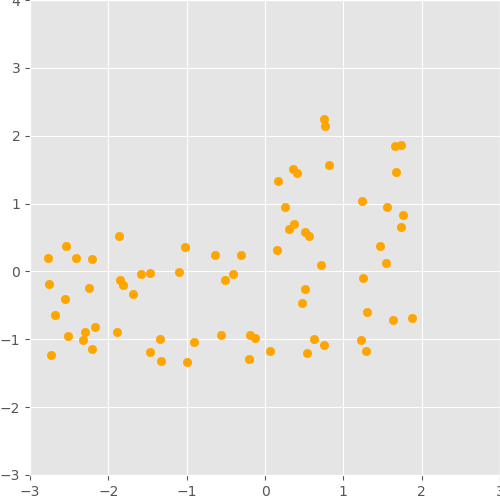}
		\end{minipage}
	\end{center}
	\caption{Examples of metric spaces constructed by sampling and randomizing plane curves of Figure \ref{fig_plane_curves}.}
	\label{fig_plane_curves_randomized}
\end{figure}

\begin{figure}
	\begin{center}
		\begin{minipage}{0.99\textwidth}
			\includegraphics[width=0.142\textwidth]{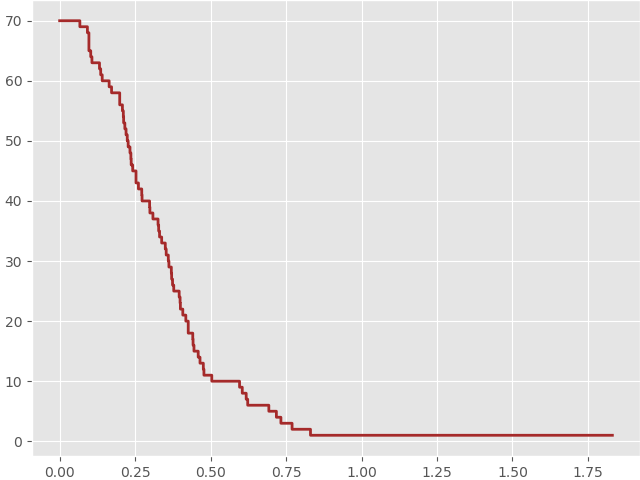}
			\hfill
			\includegraphics[width=0.142\textwidth]{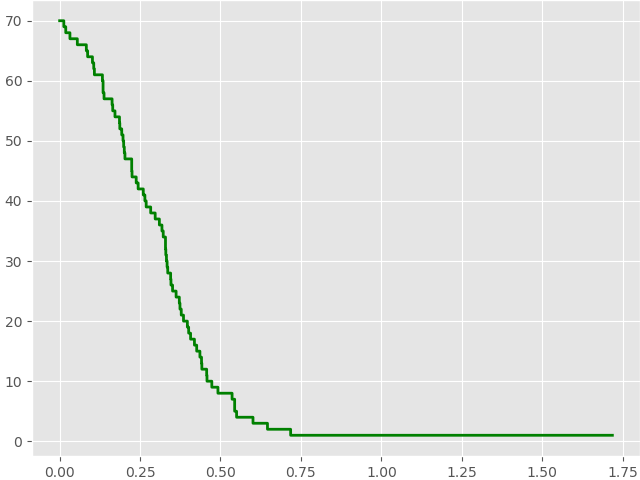}
			\hfill
			\includegraphics[width=0.142\textwidth]{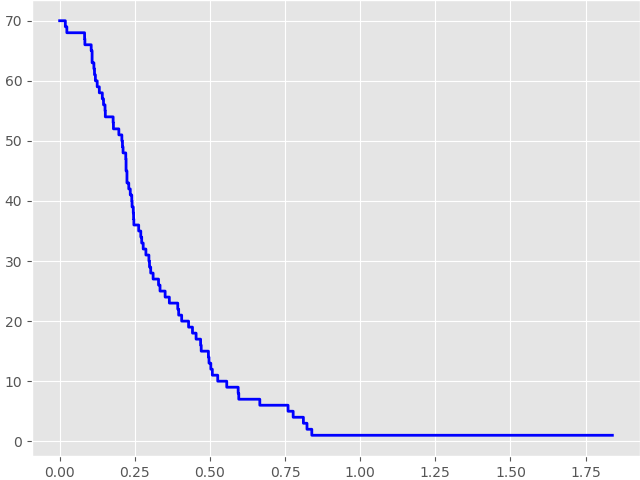}
			\hfill
			\includegraphics[width=0.142\textwidth]{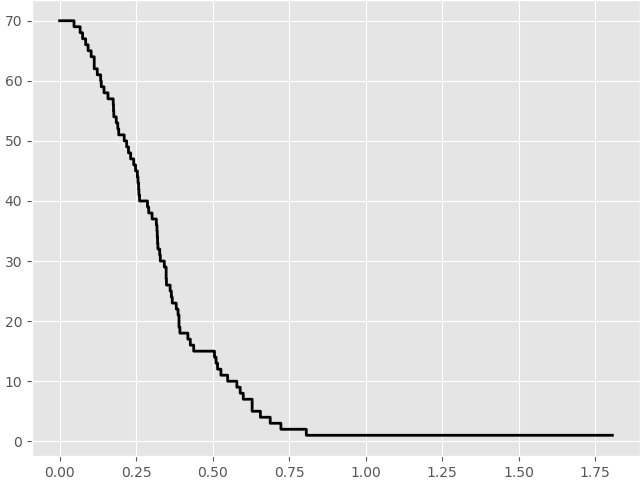}
			\hfill
			\includegraphics[width=0.142\textwidth]{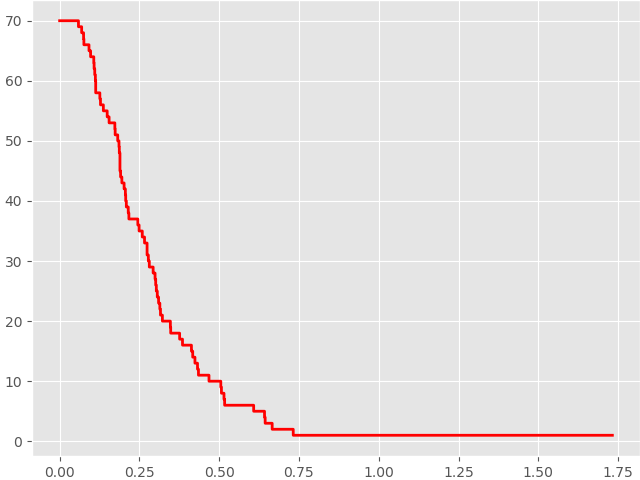}
			\hfill
			\includegraphics[width=0.142\textwidth]{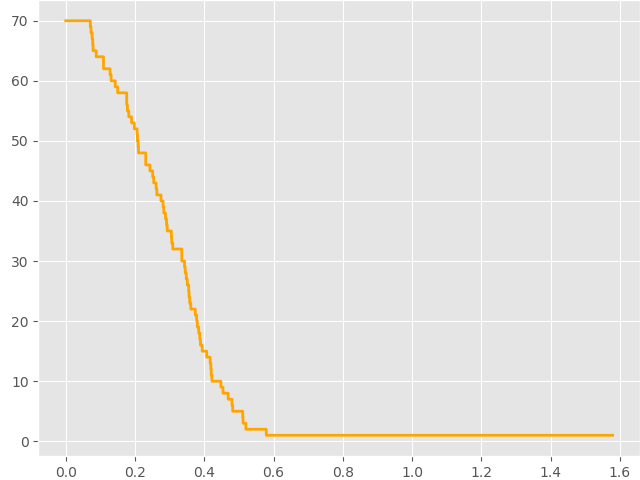}
		\end{minipage}
	\end{center}
	\begin{center}
		\begin{minipage}{0.99\textwidth}
			\includegraphics[width=0.142\textwidth]{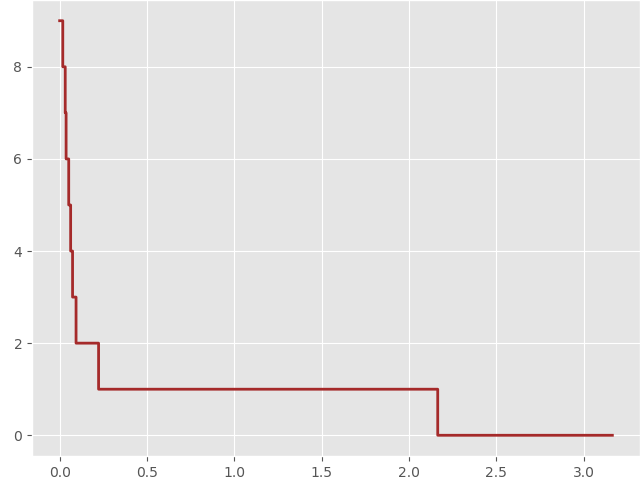}
			\hfill
			\includegraphics[width=0.142\textwidth]{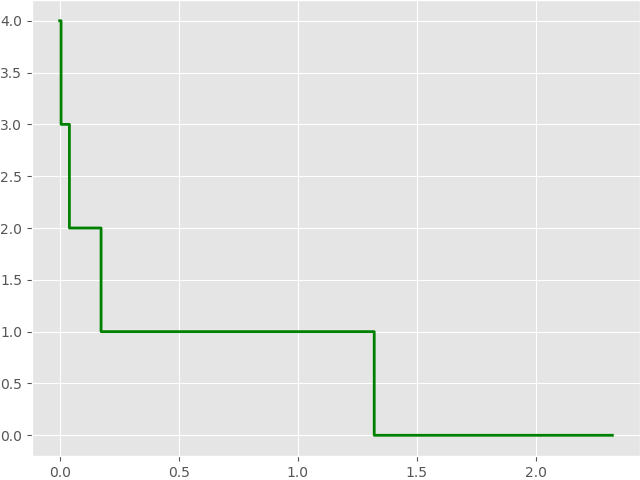}
			\hfill
			\includegraphics[width=0.142\textwidth]{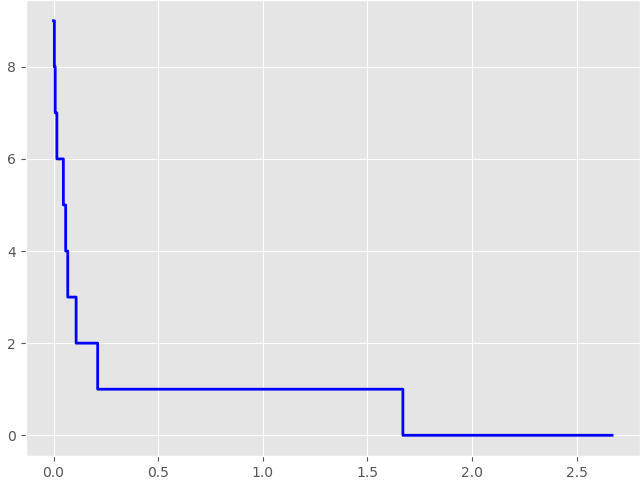}
			\hfill
			\includegraphics[width=0.142\textwidth]{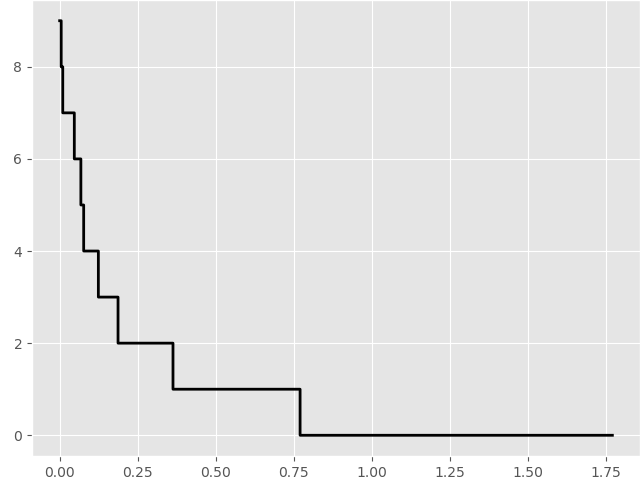}
			\hfill
			\includegraphics[width=0.142\textwidth]{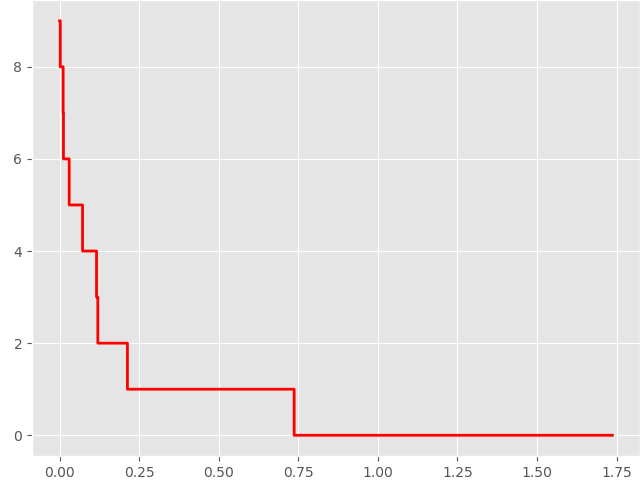}
			\hfill
			\includegraphics[width=0.142\textwidth]{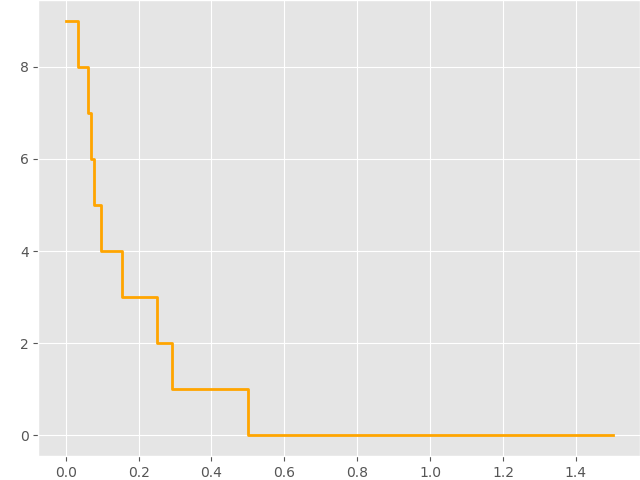}
		\end{minipage}
	\end{center}
	\caption{Stable ranks for the metric spaces in Figure \ref{fig_plane_curves_randomized}, first row for $H_0$, second row for $H_1$.}
	\label{fig_plane_curves_stable_ranks}
\end{figure}

Can the  stable ranks be used to distinguish  these curves and how can this be validated? Our approach was to  associate the following metric spaces with the curves.
We sampled 70 points uniformly distributed along the standard parameterizations of the curves. Bivariate normal distributions centered on the points were used for adding error. The distributions had diagonal covariance $\bigl[\begin{smallmatrix} \sigma^2 & 0 \\ 0 & \sigma^2 \end{smallmatrix} \bigr]$ with $\sigma=0.25$. We regard these samplings as metric spaces with the Euclidean distance. Examples for each  curve are shown in Figure \ref{fig_plane_curves_randomized}. The corresponding stable ranks $\widehat{\text{rank}}_0$ and  $\widehat{\text{rank}}_1$  are illustrated in Figure \ref{fig_plane_curves_stable_ranks} ($H_0$ in the first row and $H_1$ in the second row).

\begin{figure}
	\begin{center}
		\begin{minipage}{0.99\textwidth}
			\includegraphics[width=0.32\textwidth]{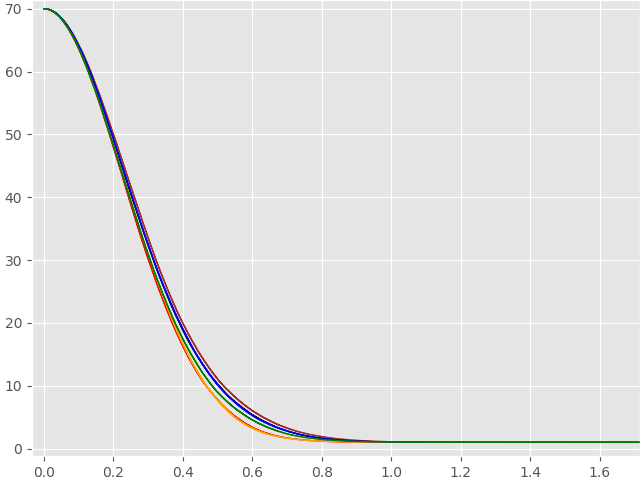}
			\hfill
			\includegraphics[width=0.32\textwidth]{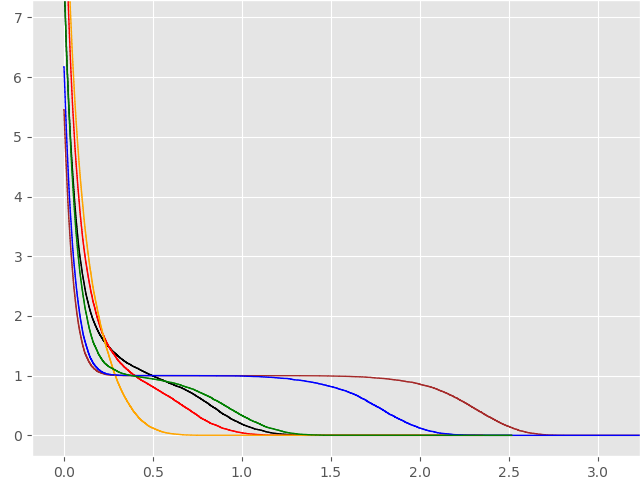}
			\hfill
			\includegraphics[width=0.32\textwidth]{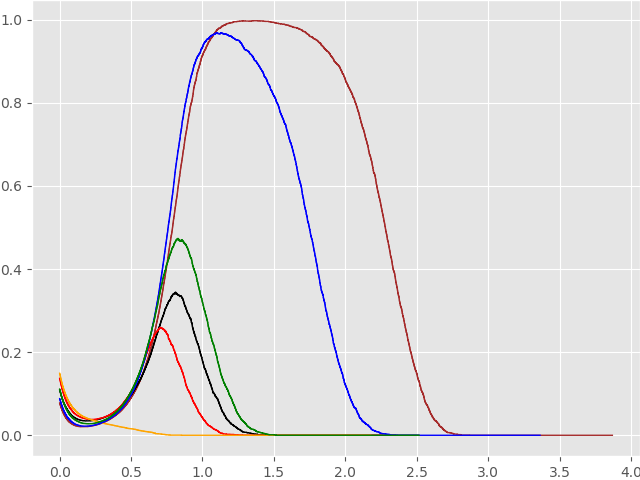}
		\end{minipage}
	\end{center}
	\caption{Graphs of the average $E(\widehat{\text{rank}}_0)$ of 2500 stable ranks $\widehat{\text{rank}}_0$ (left), the average $E(\widehat{\text{rank}}_1)$ of 2500 stable ranks $\widehat{\text{rank}}_1$ (middle) and the quotient $E(\widehat{\text{rank}}_1)/E(\widehat{\text{rank}}_0)$ (right).} 
	\label{fig_plane_curves_stable_ranks_averages}
\end{figure}

The key step  is to take advantage of the possibility of forming  averages of stable ranks which should  converge to some   expected values (the law of large numbers). The  hope is that these  expected values  are different for different curves. To test this   we repeat the above steps of choosing 70 points, forming metric spaces with the Euclidean distance and taking their stable ranks $\widehat{\text{rank}}_0$ and $\widehat{\text{rank}}_1$ 2500 times. To each curve we can then associate three functions: the pointwise average $E(\widehat{\text{rank}}_0)$ of the 2500 stable ranks $\widehat{\text{rank}}_0$, the pointwise average $E(\widehat{\text{rank}}_1)$ of the 2500 stable ranks $\widehat{\text{rank}}_1$ and the quotient $E(\widehat{\text{rank}}_1)/E(\widehat{\text{rank}}_0)$. The graphs of these functions are shown in Figure \ref{fig_plane_curves_stable_ranks_averages}.

We think about these functions as summaries of some aspects of the geometry of the plane curves. To estimate how representative these summaries are of the curves we computed the following distance tables. The  tables in Table \ref{fig_plane_curves_averages_distances} contain the pairwise integral distances between the above summaries rounded to 2 decimal points.

\begin{table}
	\begin{center}
		\scalebox{0.6}{\begin{tabular}{c|c|c|c|c|c|c}
				&{\color{brown}$\bullet$}& {\color{green}$\bullet$}& {\color{blue}$\bullet$}& {\color{black}$\bullet$}
				& {\color{red}$\bullet$}& {\color{orange}$\bullet$}\\ \hline
				{\color{brown}$\bullet$} & 0. &  1.27 &0.55& 0.59 &1.83 &1.69 \\ \hline
				{\color{green}$\bullet$} &1.27 &0.  & 0.72 &0.68 &0.56 &0.5  \\ \hline
				{\color{blue}$\bullet$}&0.55 &0.72 &0.  & 0.07 &1.28 &1.14 \\ \hline
				{\color{black}$\bullet$} &0.59 &0.68 &0.07 &0.   &1.24 &1.1  \\ \hline
				{\color{red}$\bullet$}& 1.83& 0.56 &1.28 &1.24 &0.  & 0.19 \\ \hline
				{\color{orange}$\bullet$}& 1.69 &0.5 & 1.14 &1.1 & 0.19 &0.
		\end{tabular}}\ \ \ 
		\scalebox{0.6}{\begin{tabular}{c|c|c|c|c|c|c}
				&{\color{brown}$\bullet$}& {\color{green}$\bullet$}& {\color{blue}$\bullet$}& {\color{black}$\bullet$}
				& {\color{red}$\bullet$}& {\color{orange}$\bullet$}\\ \hline
				{\color{brown}$\bullet$} & 0.  & 1.59 &0.59 &1.76 &2.05 &2.42 \\ \hline
				{\color{green}$\bullet$} &1.59 &0.  & 0.99 &0.18 &0.47 &0.84 \\ \hline
				{\color{blue}$\bullet$}& 0.59 & 0.99 & 0.  & 1.16 &1.46 &1.83 \\ \hline
				{\color{black}$\bullet$} &1.76 &0.18 &1.16 &0.   &0.33 &0.75 \\ \hline
				{\color{red}$\bullet$}& 2.05 &0.47& 1.46 &0.33 &0.   &0.42 \\ \hline
				{\color{orange}$\bullet$}& 2.42 &0.84 &1.83 &0.75 &0.42 & 0.
		\end{tabular}}\ \ \ 
		\scalebox{0.6}{\begin{tabular}{c|c|c|c|c|c|c}
				&{\color{brown}$\bullet$}& {\color{green}$\bullet$}& {\color{blue}$\bullet$}& {\color{black}$\bullet$}
				& {\color{red}$\bullet$}& {\color{orange}$\bullet$}\\ \hline
				{\color{brown}$\bullet$} & 0. &  1.28 &0.56 &1.34 &1.4 & 1.49 \\ \hline
				{\color{green}$\bullet$} &1.28 &0.  & 0.76 &0.07 &0.13 & 0.22 \\ \hline
				{\color{blue}$\bullet$}&0.56& 0.76 &0. &  0.83 & 0.89 &0.98 \\ \hline
				{\color{black}$\bullet$} &1.34 &0.07 &0.83 &0.   &0.06 &0.15 \\ \hline
				{\color{red}$\bullet$}& 1.4 & 0.13 &0.89 &0.06 &0.  & 0.1\\ \hline
				{\color{orange}$\bullet$}& 1.49 &0.22 &0.98 &0.15 &0.1  &0.
		\end{tabular}}
	\end{center}
	\caption{Pairwise integral distances of the summaries in Figure \ref{fig_plane_curves_stable_ranks_averages}: left $E(\widehat{\text{rank}}_0)$, middle 
		$E(\widehat{\text{rank}}_1)$, and right $E(\widehat{\text{rank}}_1)/E(\widehat{\text{rank}}_0)$.}
	%
	\label{fig_plane_curves_averages_distances} 
\end{table}

We performed the  steps described above of taking the summaries of 2500 samplings of the 6 curves 10 times. Every time  we obtain tables of the integral distances between the  summaries. We then ask how these  distances vary. The tables in Table \ref{fig_plane_curves_averages_distances_variation} contain the differences between the maximum and the minimum value at each entry of these 10 distance matrices rounded to 3 decimal points.

\begin{table}
	\begin{center}
		\scalebox{0.54}{\begin{tabular}{c|c|c|c|c|c|c}
				&{\color{brown}$\bullet$}& {\color{green}$\bullet$}& {\color{blue}$\bullet$}& {\color{black}$\bullet$}
				& {\color{red}$\bullet$}& {\color{orange}$\bullet$}\\ \hline
				{\color{brown}$\bullet$} & 0.  &   0.052  &0.089  &0.053&  0.09  & 0.034 \\ \hline
				{\color{green}$\bullet$} &0.052 & 0.  &   0.079 & 0.057 & 0.08 &  0.057 \\ \hline
				{\color{blue}$\bullet$}&0.089 & 0.079 & 0.   &  0.045 & 0.062  &0.076 \\ \hline
				{\color{black}$\bullet$} &0.053 & 0.057  & 0.045 & 0.  &   0.071 & 0.062  \\ \hline
				{\color{red}$\bullet$}& 0.09 &  0.08 &  0.062 & 0.071 & 0.  &   0.127 \\ \hline
				{\color{orange}$\bullet$}& 0.034 & 0.057 & 0.076 & 0.062  & 0.127 & 0. 
		\end{tabular}}\ \ 
		\scalebox{0.54}{\begin{tabular}{c|c|c|c|c|c|c}
				&{\color{brown}$\bullet$}& {\color{green}$\bullet$}& {\color{blue}$\bullet$}& {\color{black}$\bullet$}
				& {\color{red}$\bullet$}& {\color{orange}$\bullet$}\\ \hline
				{\color{brown}$\bullet$} & 0.  &   0.022 & 0.024 & 0.014 & 0.032 & 0.021\\ \hline
				{\color{green}$\bullet$} &0.022 & 0.   &  0.027 & 0.016 & 0.02 &  0.017 \\ \hline
				{\color{blue}$\bullet$}& 0.024 & 0.027 & 0.  &   0.026 & 0.029 & 0.032 \\ \hline
				{\color{black}$\bullet$} &0.014 & 0.016 & 0.026 & 0.  &   0.028 & 0.025\\ \hline
				{\color{red}$\bullet$}& 0.032 & 0.02  & 0.029 & 0.028 & 0.   &  0.028 \\ \hline
				{\color{orange}$\bullet$}& 0.021 & 0.017 & 0.032 & 0.025 & 0.028 & 0. 
		\end{tabular}}\ \ 
		\scalebox{0.54}{\begin{tabular}{c|c|c|c|c|c|c}
				&{\color{brown}$\bullet$}& {\color{green}$\bullet$}& {\color{blue}$\bullet$}& {\color{black}$\bullet$}
				& {\color{red}$\bullet$}& {\color{orange}$\bullet$}\\ \hline
				{\color{brown}$\bullet$} & 0. &    0.017 & 0.015 & 0.019 & 0.022 & 0.017 \\ \hline
				{\color{green}$\bullet$} & 0.017 & 0.   &  0.021 & 0.009 & 0.011 & 0.008\\ \hline
				{\color{blue}$\bullet$}&0.015 & 0.021 & 0.  &   0.02  & 0.019 & 0.018 \\ \hline
				{\color{black}$\bullet$} &0.019 & 0.009 & 0.02  & 0.  &   0.004 & 0.003  \\ \hline
				{\color{red}$\bullet$}& 0.022 & 0.011&  0.019 & 0.004 & 0.   &  0.01 \\ \hline
				{\color{orange}$\bullet$}& 0.017 & 0.008 & 0.018&  0.003 & 0.01 &  0.   
		\end{tabular}}
	\end{center}
	\caption{Absolute variation of the integral distances in tables of Table \ref{fig_plane_curves_averages_distances} after 10 repeats for $E(\widehat{\text{rank}}_0)$(left), $E(\widehat{\text{rank}}_1)$ (middle), and
		$E(\widehat{\text{rank}}_1)/E(\widehat{\text{rank}}_0)$ (right).}
	\label{fig_plane_curves_averages_distances_variation} 
\end{table}

These differences are small compared to the distances described above, variation not exceeding 10\%. We can conclude that the summaries do indeed distinguish different curves in a stable manner when the curves are subjected to random sampling and error. The distance variations are particularly small
for the third summary $E(\widehat{\text{rank}}_1)/E(\widehat{\text{rank}}_0)$. It is interesting to note that combining $H_0$ and $H_1$ summaries in this way might be more stable or more informative of the data than only $H_0$ clustering or $H_0$ and $H_1$ separately.

\begin{figure}
	\begin{center}
		\begin{minipage}{\textwidth}
			\includegraphics[width=0.99\textwidth]{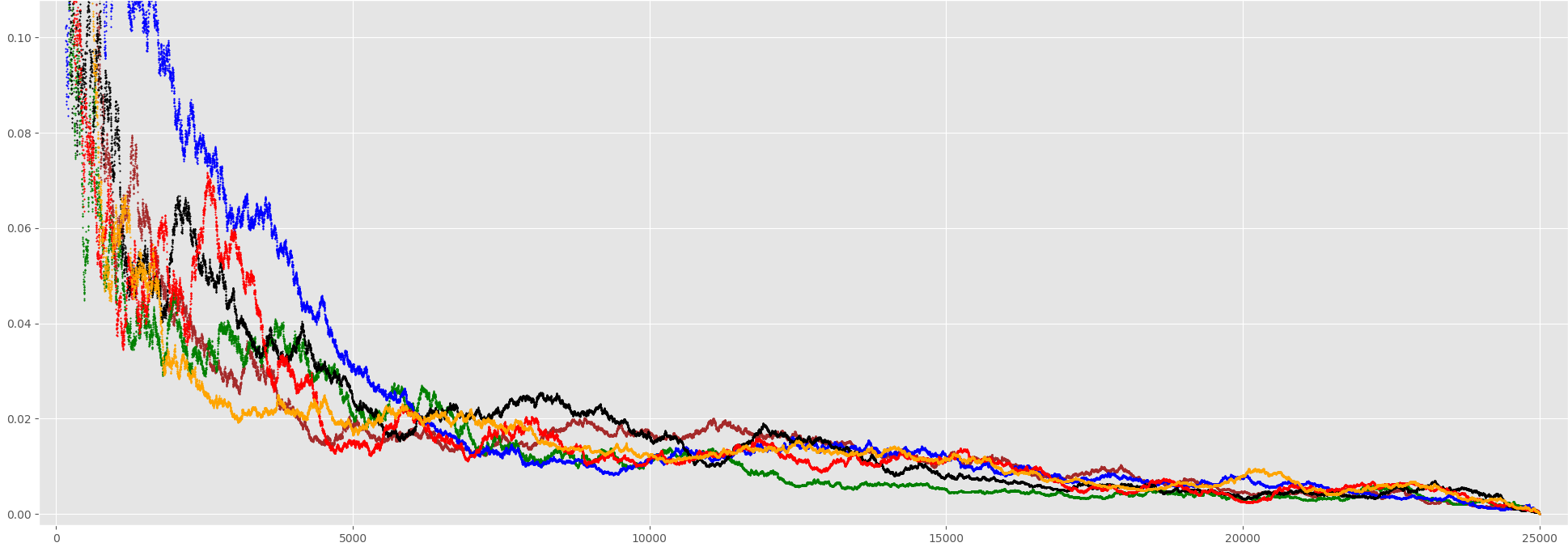}
		\end{minipage}
	\end{center}
	\begin{center}
		\begin{minipage}{\textwidth}
			\includegraphics[width=0.99\textwidth]{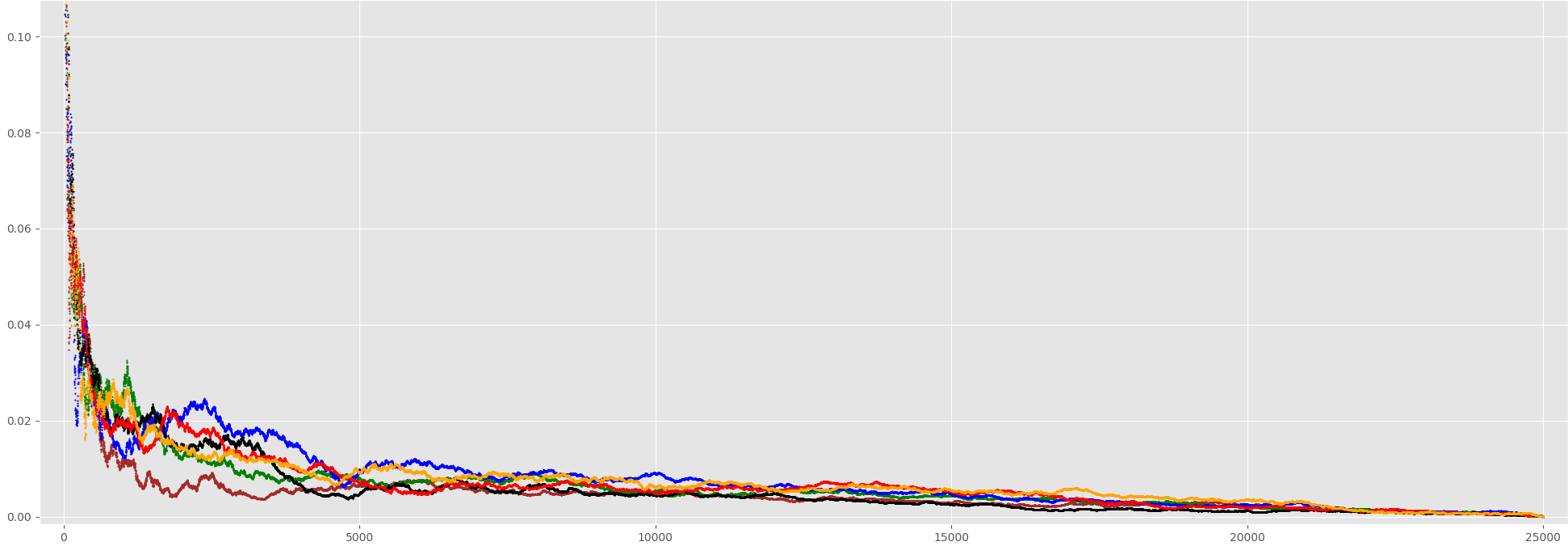}
		\end{minipage}
	\end{center}
	\caption{$d_{\int}(E_{n,0},E_{25000,0})$ (top) and $d_{\int}(E_{n,1},E_{25000,1})$ (bottom) for $1\leq n\leq 25000$.}
	\label{fig_plane_curves_central_limit}
\end{figure}

To illustrate the law of large numbers and the central limit behaviour we considered, for each curve, 25000 samplings of 70 points as described at the beginning of this section. For  $n$ in $1\leq n\leq 25000$, we took the averages of the first $n$ stable ranks,  denoting them by $E_{n,0}=E_n(\widehat{\text{rank}}_0)$ and $E_{n,1}=E_n(\widehat{\text{rank}}_1)$, and then  we computed the distances 
$d_{\int}(E_{n,0},E_{25000,0})$ 
and
$d_{\int}(E_{n,1},E_{25000,1})$. 
Figure \ref{fig_plane_curves_central_limit} is a plot of these distances as functions of $n$. 
These plots clearly show convergence of the means. Persistent homology is very well known to be stable (\cite{algebraic_stability},\cite{original_stability}). However, even single points can  generate unstable $H_0$ homology classes in  noisy data. One way to stabilize  a persistence analysis is to make repeated samplings of the data and take the average summaries. We have aimed to demonstrate the usefulness of stable ranks in this regard. Another point to make is that higher-dimensional homological information is very useful, recall Figure \ref{fig_plane_curves_stable_ranks_averages}. Figure \ref{fig_plane_curves_central_limit} illustrates this further from statistical point of view: convergence of the means is clearly slower for $H_0$ than for $H_1$.

\section{Using contours}\label{contour_usage}
In this section we illustrate how contours and the induced  stable ranks can be used in supervised learning. We emphasize that focus is not on finding optimal classifier for a specific case but rather to demonstrate how one uses our pipeline, particularly for feature extraction. Our classifiers also rely on taking averages which we are able to do with stable ranks. Two case studies are considered, point processes on a unit square and real data from human activities.

\subsection{Point processes}\label{point_processes}
Point processes have gathered interest in TDA community, see for example \cite{ APF, LimitTheorems, HypothesisTesting}. We simulated six different classes of point processes on a unit square, see their descriptions below. For each class we produced 500 simulations on average containing 200 points. Let $X \sim PD(k)$ denote that random variable $X$ follows probability disribution $PD$ with parameter $k$. In particular, $\text{Poisson}(\lambda)$ denotes the Poisson distribution with event rate $\lambda$.
\smallskip

\noindent
\textbf{Poisson}: \quad We first sampled number of events $N$, where $N \sim \text{Poisson}(\lambda)$. We then sampled $N$ points from a uniform distribution defined on the unit square $[0,1] \times [0,1]$. Here $\lambda=200$.
\smallskip

\noindent
\textbf{Normal}: \quad Again number of events $N$ was sampled from $\text{Poisson}(\lambda)$, $\lambda = 200$. We then created $N$ coordinate pairs $(x,y)$, where both $x$ and $y$ are sampled from normal distribution $N(\mu,\sigma^2)$ with mean $\mu$ and standard deviation $\sigma$. Here $\mu=0.5$ and $\sigma = 0.2$.
\smallskip

\noindent
\textbf{Matern}: \quad Poisson process as above was simulated with event rate $\kappa$. Obtained points represent parent points, or cluster centers, on the unit square. For each parent, number of child points $N$ was sampled from $\text{Poisson}(\mu)$. A disk of radius $r$ centered on each parent point was defined. Then for each parent the corresponding number of child points $N$ were placed on the disk. Child points were uniformly distributed on the disks. Note that parent points are not part of the actual data set. We set $\kappa$=40, $\mu$=5 and $r=0.1$.
\smallskip

\noindent
\textbf{Thomas}: \quad Thomas process is similar to Matern process except that instead of uniform distributions, child points were sampled from bivariate normal distributions defined on the disks. The distributions were centered on the parents and had diagonal covariance $\bigl[\begin{smallmatrix} \sigma^2 & 0 \\ 0 & \sigma^2 \end{smallmatrix} \bigr]$. Here $\sigma=0.1$. 
\smallskip

\noindent
\textbf{Baddeley-Silverman}: \quad For this process the unit square was divided into equal size squares with side lengths $\frac{1}{14}$. Then for each tile number of points $N$ was sampled, $N \sim \text{Baddeley-Silverman}$. Baddeley-Silverman distribution is a discrete distribution defined on values $(0,1,10)$ with  probabilities  $(\frac{1}{10},\frac{8}{9},\frac{1}{90})$. For each tile, associated number of points $N$ were then uniformly distributed on the tile.
\smallskip

\noindent
\textbf{Iterated function system (IFS)}:\quad  We also generated point sets with an iterated function system. For this a discrete distribution is defined on values $(0,1,2,3,4)$ with corresponding probabilities 
$\left(\frac{1}{3},\frac{1}{6},\frac{1}{6},\frac{1}{6},\frac{1}{6}\right)$. We denote this distribution by IFS. Number of points $N$ was then sampled,  $N \sim \text{Poisson}(\lambda)$, $\lambda = 200$. Starting from an initial point $(x_0,y_0)$ on the unit square, $N$ new points are generated by the recursive formula $(x_n,y_n) = f_i(x_{n-1},y_{n-1}),$ where $n \in \{1,...,N\}$, $i \sim \text{IFS}$ and the functions $f_i$ are given as
\[f_0(y,x) = \left(\frac{x}{2},\frac{y}{2}\right), f_1(y,x)= \left(\frac{x}{2}+\frac{1}{2},\frac{y}{2}\right), 
f_2(y,x) = \left(\frac{x}{2},\frac{y}{2}+\frac{1}{2}\right)\] 
\[f_3(y,x) = \left(\left|\frac{x}{2}-1\right|,\frac{y}{2}\right), 
f_4(y,x)= \left(\frac{x}{2},\left|\frac{y}{2}-1\right|\right). 
\]
\smallskip

\begin{figure}[!h]
	\begin{minipage}{0.99\textwidth}
		\includegraphics[width=0.49\textwidth]{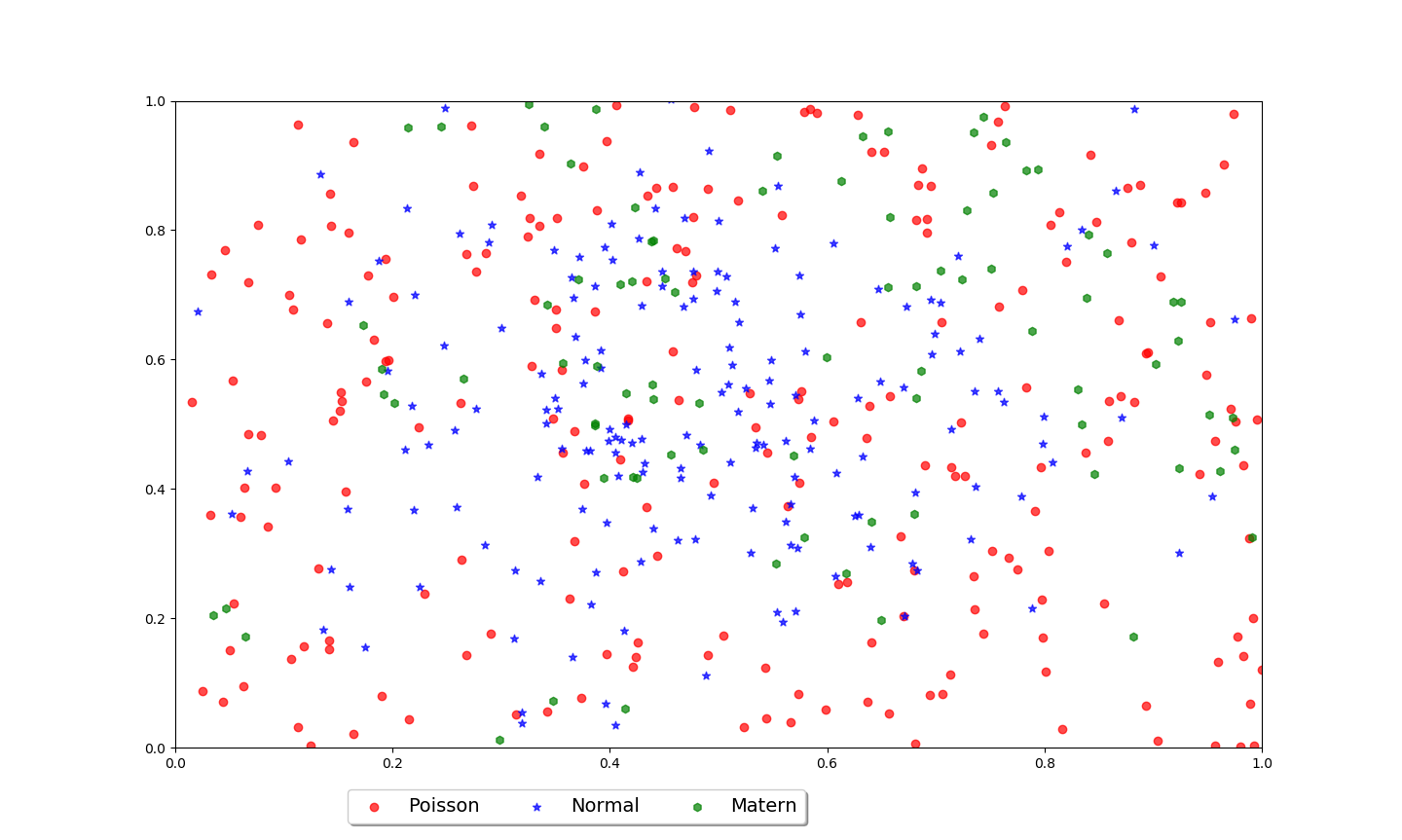}
		\includegraphics[width=0.49\textwidth]{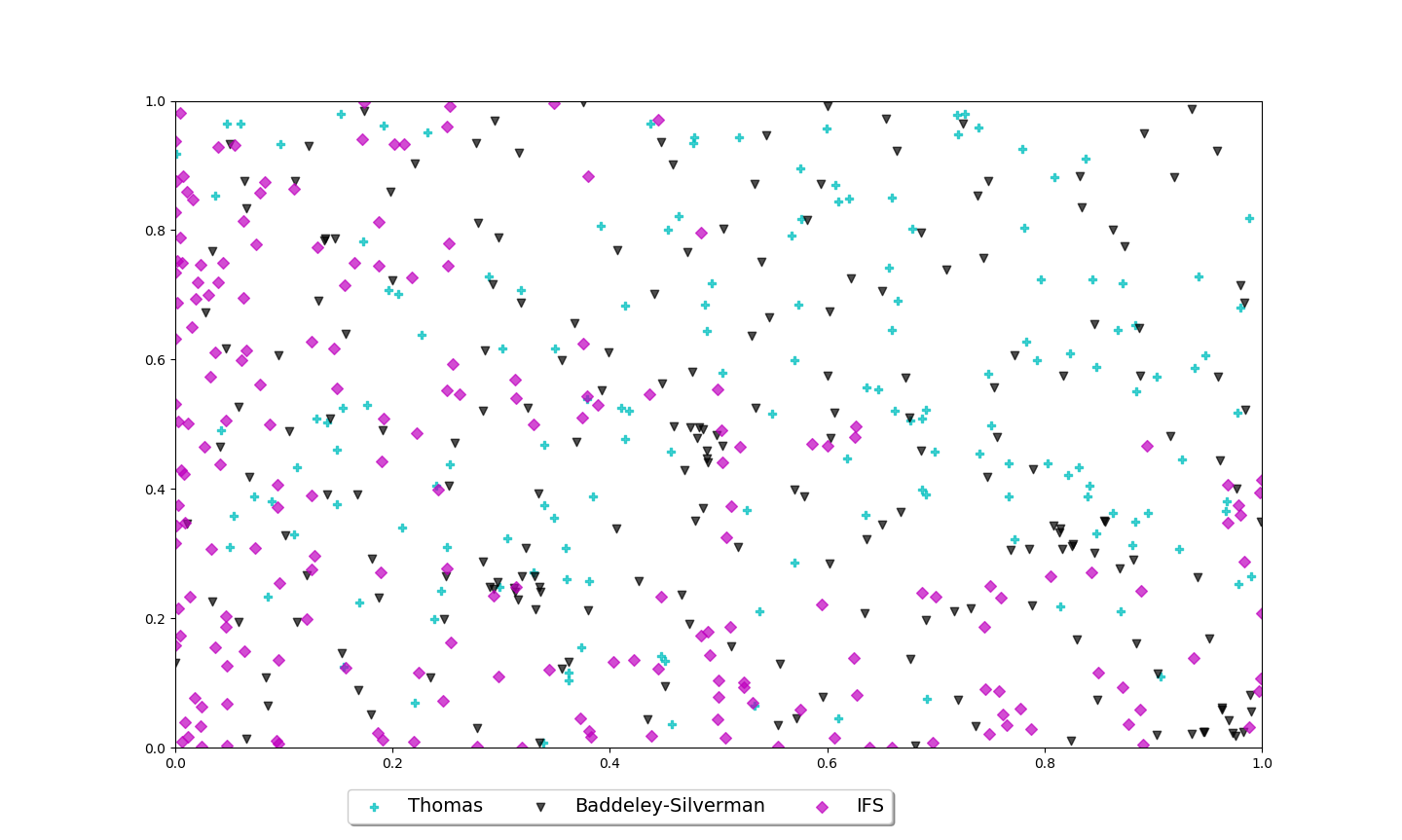}	
	\end{minipage}
	\caption{Example realizations of point processes on unit square.}
	\label{fig_point_processes}
\end{figure}

Figure \ref{fig_point_processes} shows one realization of the point processes with given parameters. From topological data analysis point of view the point sets hold no distinct large scale topology. It is therefore ideal to study the geometric correlations or features in the filtration captured by homologies in degrees 0 and 1. 
The number of these features on different scales are captured by stable ranks   as demonstrated in the rest of this section.

\begin{figure}[!h]
	\begin{minipage}{0.99\textwidth}
		\includegraphics[width=0.49\textwidth,clip,trim=2cm 1.7cm 2cm 2.5cm]{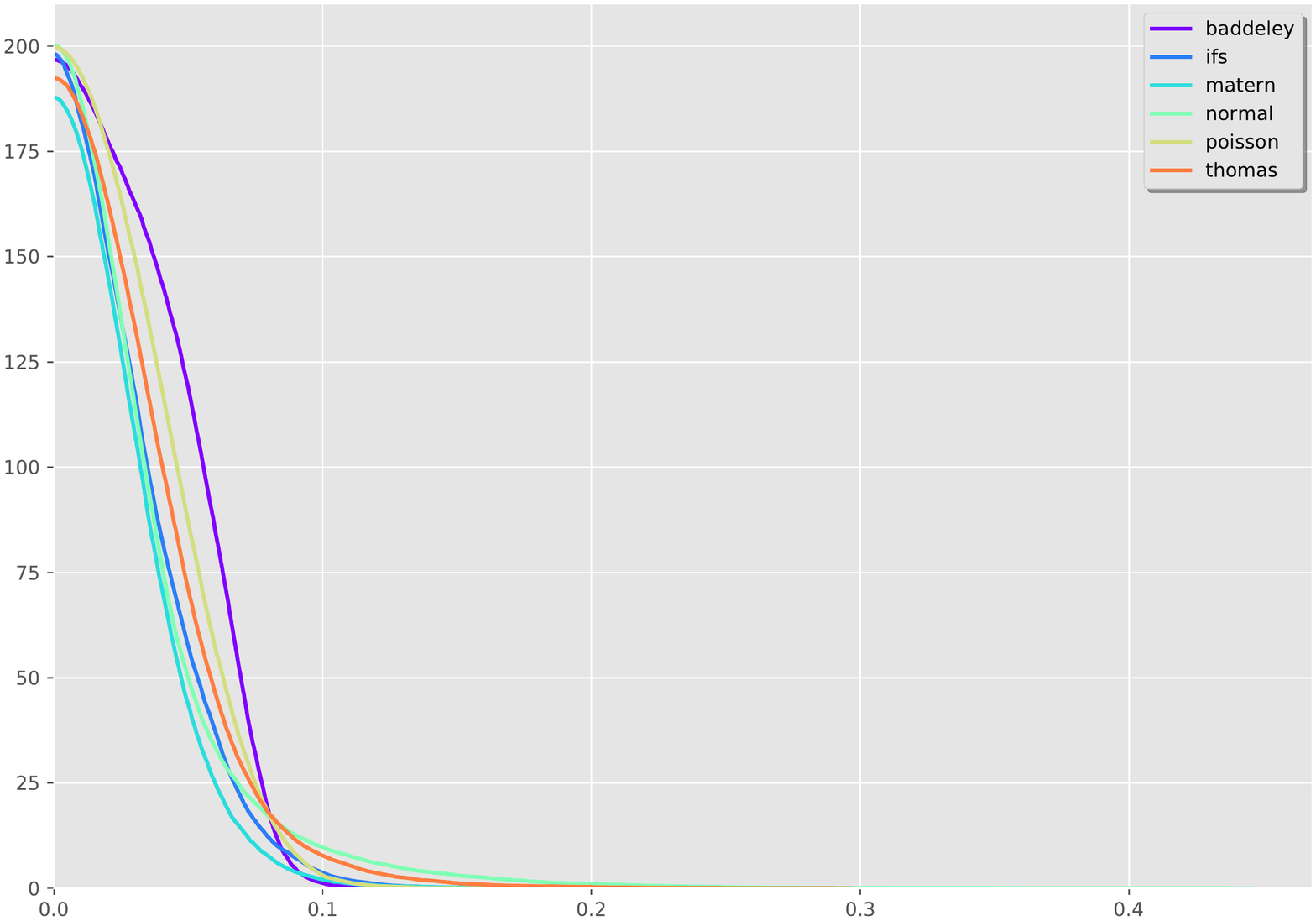}
		\includegraphics[width=0.49\textwidth,clip,trim=2cm 1.7cm 2cm 2.5cm]{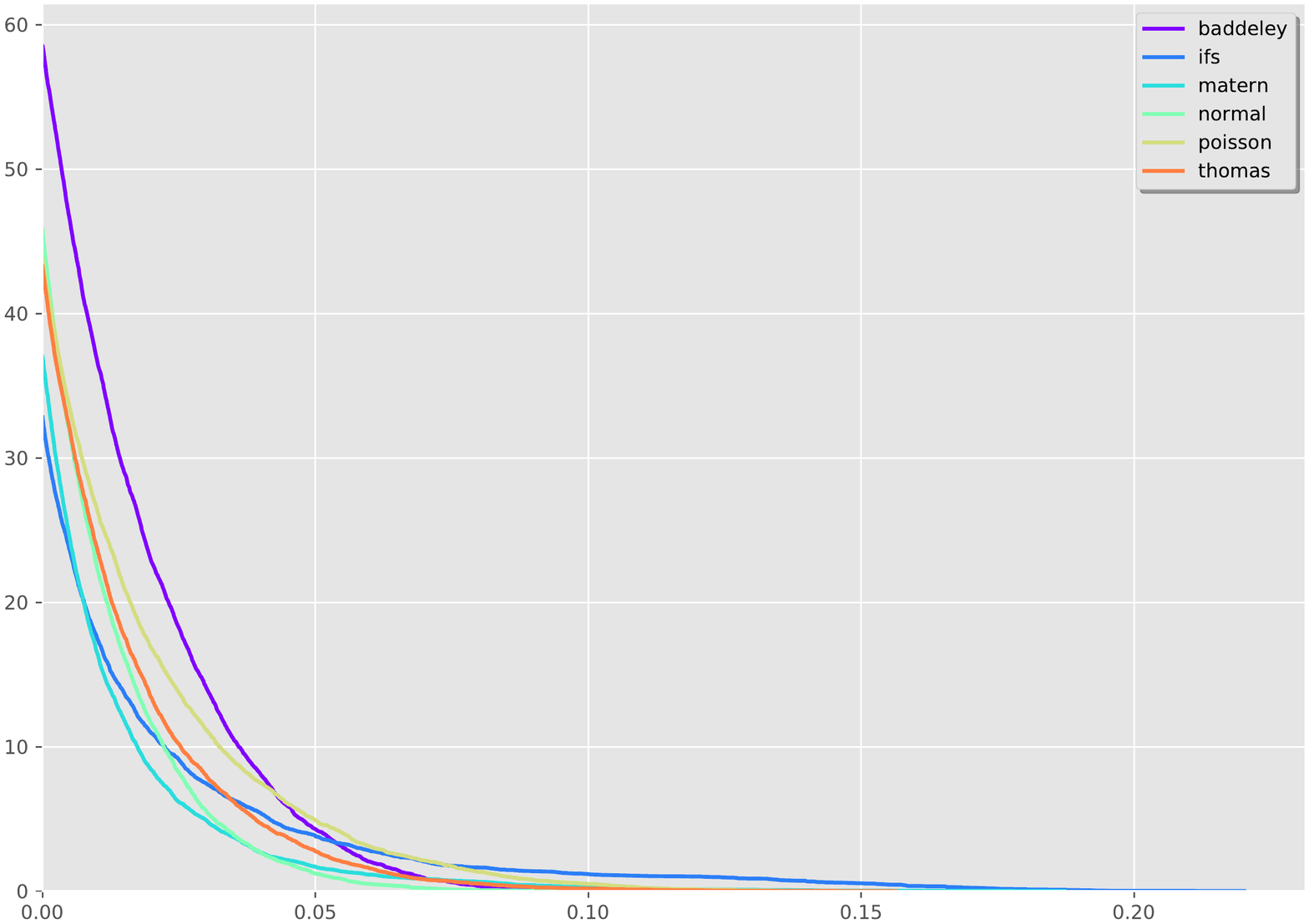}	
	\end{minipage}
	\caption{Mean stable ranks with respect to the standard contour  for $H_0$ (left) and $H_1$ (right) for 200 simulations.}
	\label{fig_point_processes_avg200_stan}
\end{figure}

Figure \ref{fig_point_processes_avg200_stan} is a plot of the averages (point-wise means) of $H_0$ and $H_1$ stable ranks  with respect to the standard contour  for 200 simulations of the point processes. Different point processes are clearly distinguished by their topological signatures. It is worth noting that Matern and Thomas processes are well separated even though in their definition they only differ in the distribution used for point clusters. 

To test how well the stable ranks with respect to the standard contour perform in classifying different point processes we conduced  mean classification procedure as follows. For each class we chose 200 simulations out of the 500 produced. We then computed the point-wise means of their  stable ranks with respect to the standard contour. This mean invariant was then used as a classifier. For the remaining 300 signatures in each class we computed integral distance between each signature and all classifiers. Found minimum distance was recorded by adding 1 to the corresponding pair of the classifier and the test  class. Classification is successful if the classifier and the test  belong to the same class. Thus in the optimal case the value of the pair (Poisson classifier class, Poisson test class) would be 300, for example. 

\begin{figure}
	\begin{minipage}{0.99\textwidth}
		\includegraphics[width=0.49\textwidth,clip,trim=7cm 2cm 7.3cm 1.2cm]{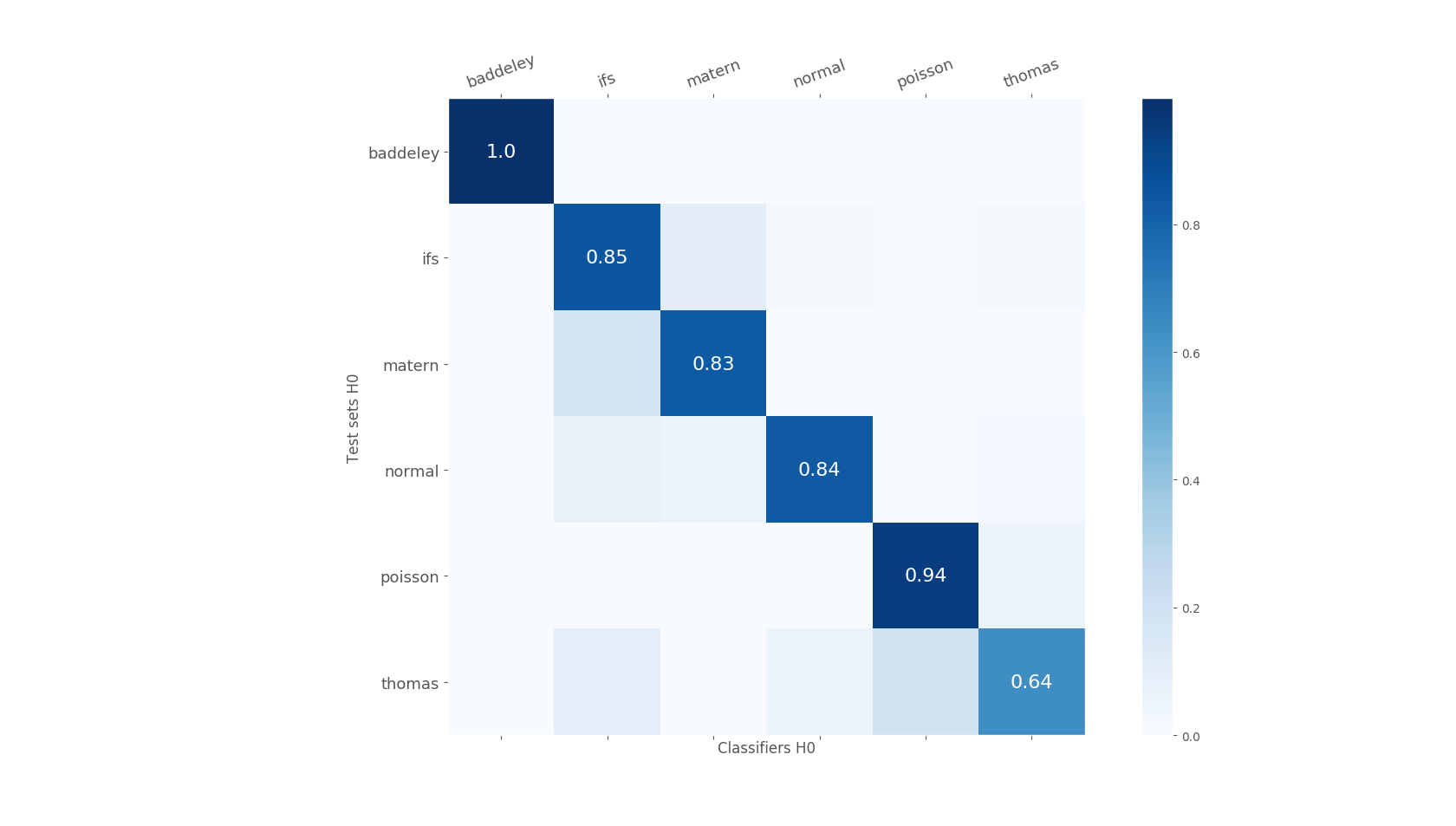}
		\includegraphics[width=0.49\textwidth,clip,trim=7cm 2cm 7.3cm 1.2cm]{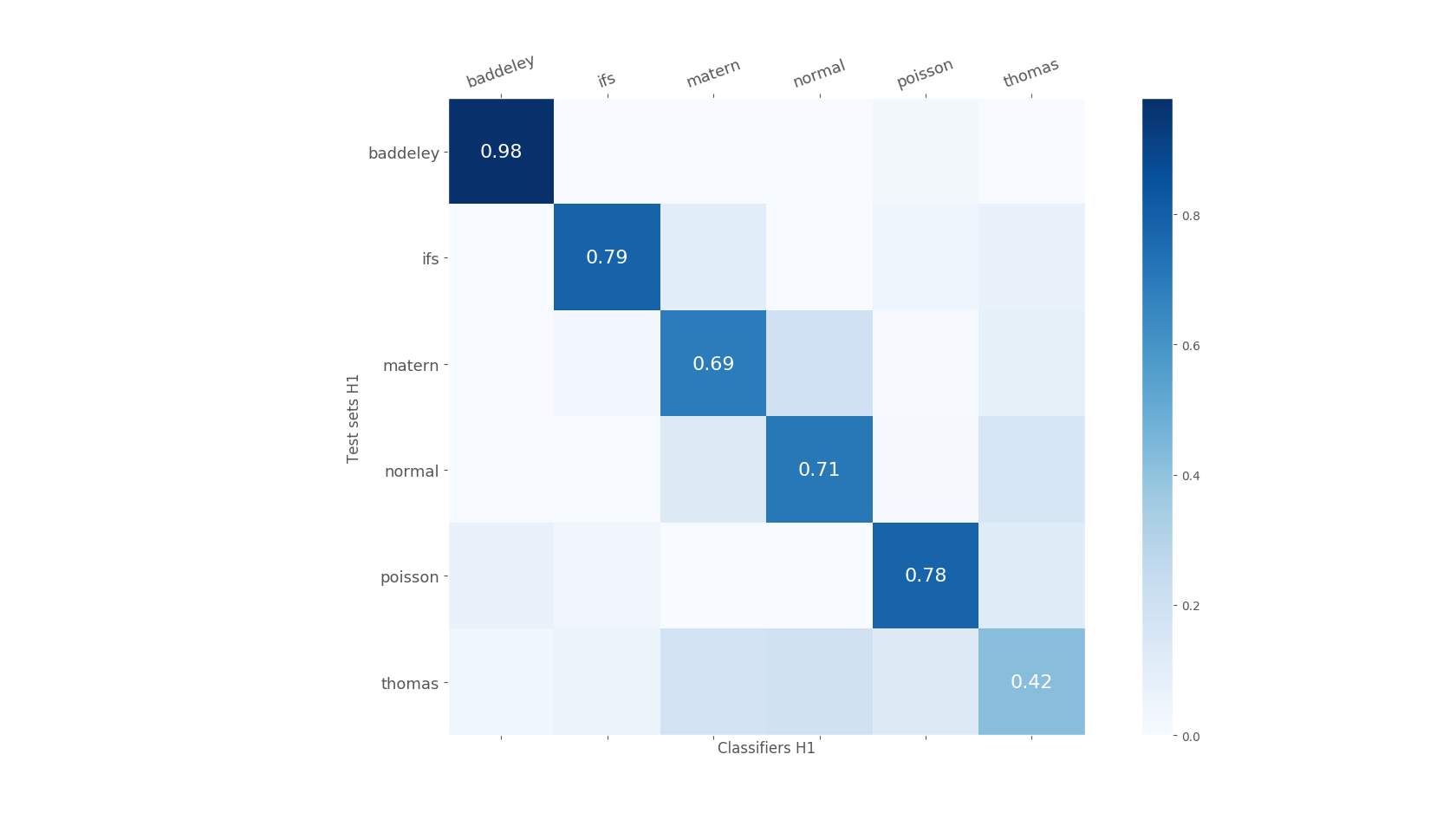}
	\end{minipage}
	\caption{Confusion matrices for the point process classification in $H_0$ (left) and $H_1$ (right).}
	\label{fig_point_proc_classification_acc_integral_stan}
\end{figure}

For cross-validation we used 20-fold random subsampling. We randomly sampled 200 stable ranks for classifiers and remaining 300 invariants in each class constituted the  test sets. This was repeated 20 times and classification accuracy was taken to be the average over the folds. Obtained cross-validated classification accuracies are reported in the confusion matrices of Figure \ref{fig_point_proc_classification_acc_integral_stan}. The confusion matrices show relative accuracies after dividing by 300 after each cross-validation fold and averaging after the full run. The mean classification accuracy by taking the average over classes (average of the diagonal) is 85\% for $H_0$ and 73\% for $H_1$. The classification procedure performs comparably or better as the hypothesis testing against the homogeneous Poisson process in \cite{APF}. Note that no other assumptions or parameter selections were involved in our methodology other than the split between training and test samples (200 and 300, respectively.)

\begin{figure}
	\begin{minipage}{0.99\textwidth}
		\centering
		\includegraphics[width=0.46\textwidth,clip,trim=0cm 0.3cm 1cm 0.5cm]{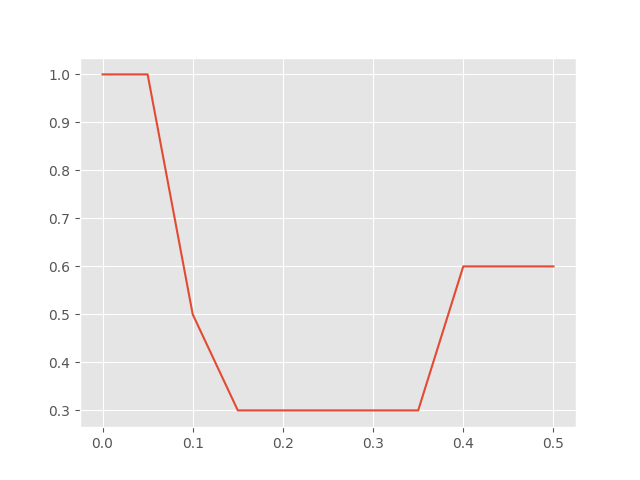}
		\includegraphics[width=0.45\textwidth,clip,trim=2cm 1cm 2.8cm 1cm]{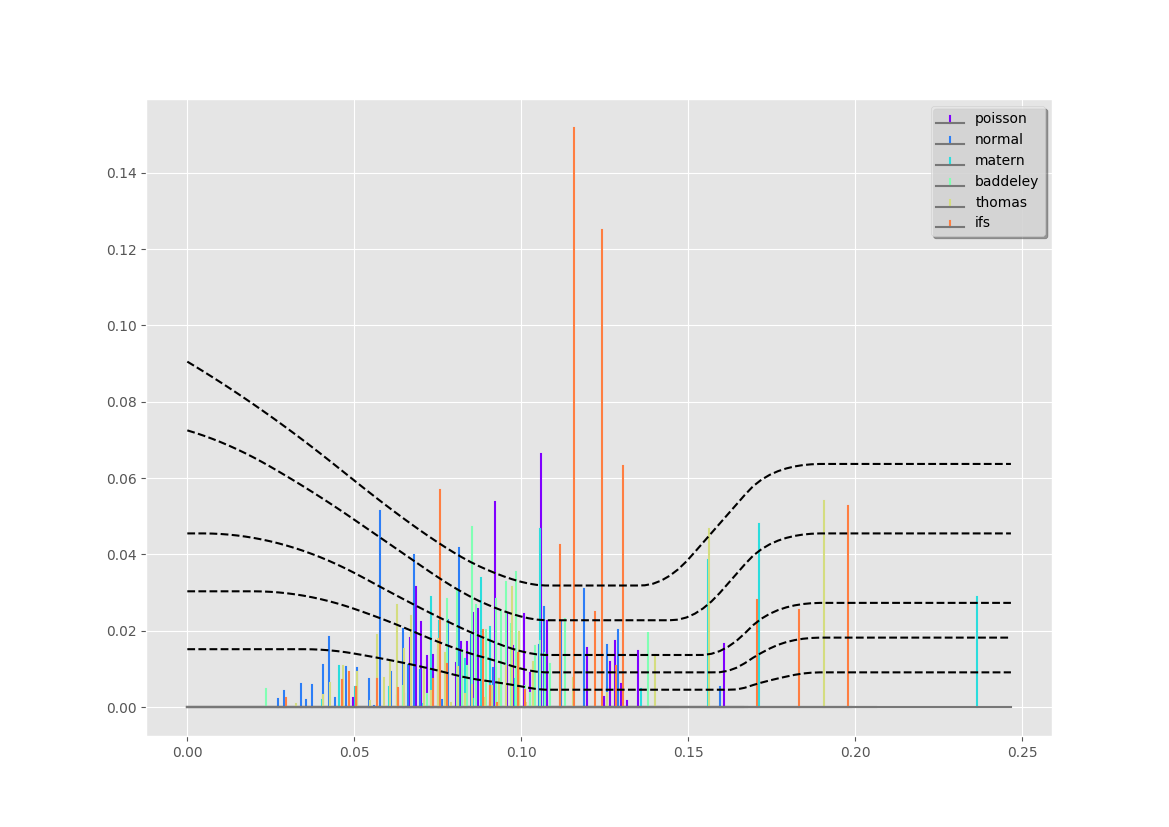}
	\end{minipage}
	\caption{Used density function in producing contour for point process classification in $H_1$ (left). Corresponding contour lines for few values of $\epsilon$ (right). Stem plot is from $H_1$ persistence analysis of one realization of the studied processes.}
	\label{fig_point_processes_contour}
\end{figure}

\begin{figure}
	\centering
	\includegraphics[scale=0.3,clip,trim=7cm 2cm 2cm 1.2cm]{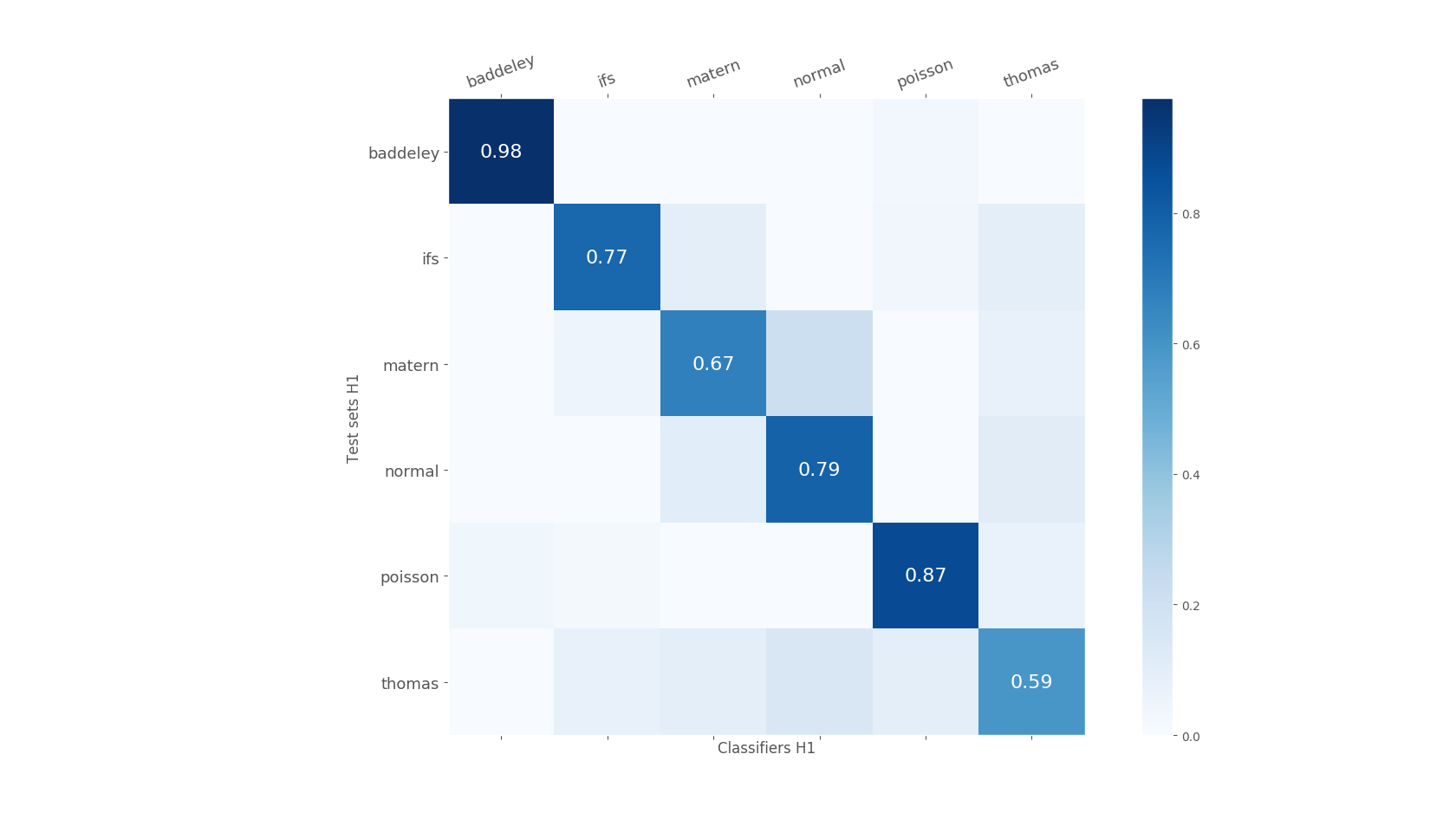}
	\caption{Confusion matrix for the classification in $H_1$ with contour coming from density function in Figure \ref{fig_point_processes_contour}.}
	\label{fig_crossval_classification_accuracies_H1_integral_contoured_200_78}
\end{figure}

To illustrate how a classification task can be improved  by using contours we performed exactly the same procedure as described above but for stable ranks of $H_1$ with respect to the shift type contour (see~\ref{shift type}) with density shown in Figure \ref{fig_point_processes_contour}. We thus put more weight on the features appearing in the middle of the filtration. In Figure \ref{fig_point_processes_contour} we also visualize contour lines for few values of $\epsilon$ as explained in Section \ref{contour_visualization}. 
The overall classification accuracy increased to 78\%, see Figure \ref{fig_crossval_classification_accuracies_H1_integral_contoured_200_78}. Particularly classification accuracy of the Thomas process was drastically improved as shown in the confusion matrix. Also noteworthy is the improvement in the accuracy of normal and Poisson processes. Contour thus captures relevant homological features from the filtration. 

\subsection{Activity monitoring}
As an application to real data we studied activity monitoring of different physical activities. Used data set was PAMAP2 data obtainable from \cite{PAMAP}. As shown in Section \ref{stability}, higher degree homology contains information for distinguishing data sets. It thus makes sense to combine homologies of different degrees into single classification scheme. In this section we demonstrate how this is enabled by our pipeline. 

Our data consisted of seven persons from the PAMAP data performing different activities such as walking, cycling, vacuuming or sitting. Test subjects were fitted with three Inertial Measurements Units (IMUs), one on wrist, ankle and chest, and a heart rate monitor. Measurements were registered every 0.1 seconds. Each IMU measured 3D acceleration, 3D gyroscopic and 3D magnetometer data. One data set thus consisted 28-dimensional data points indexed by 0.1 second timesteps. 

We chose to look at two activities in this case study: ascending and descending stairs. At the outset one would expect these activities to be very similar and therefore difficult to distinguish. For persistence analysis we randomly sampled without replacement 100 points from each data set, repeated 100 times. For each subject we thus obtained 100 resamplings from the activity data. We computed $H_0$ and $H_1$ persistence for the sampled data. The classification procedure was the same as outlined in Section \ref{point_processes} except we combined both homologies in the classifier as follows. We took the mean of 40 stable ranks both in $H_0$ and $H_1$. Classifier signatures are denoted by $\hat{P}_{H_0}$ and $\hat{P}_{H_1}$. We have 14 pairs $(\hat{P}_{H_0},\hat{P}_{H_1})$ corresponding to all (subject, activity) combinations. Remaining 60 signatures in $H_0$ and $H_1$ were used as test data and denoted $T_{H_0}$ and $T_{H_1}$. For a pair $(T_{H_0}, T_{H_1})$ from a single resampling we then found 
$$\text{min}(d_{\int}(\hat{P}_{H_0},T_{H_0}) + d_{\int}(\hat{P}_{H_1},T_{H_1})).$$
Again the classification is successful if the minimum is obtained with $\hat{P}$ and $T$ belonging to the same (subject, activity) class.

\begin{figure}
	\centering
	\includegraphics[scale=0.5,clip,trim=0cm 1cm 0cm 0cm]{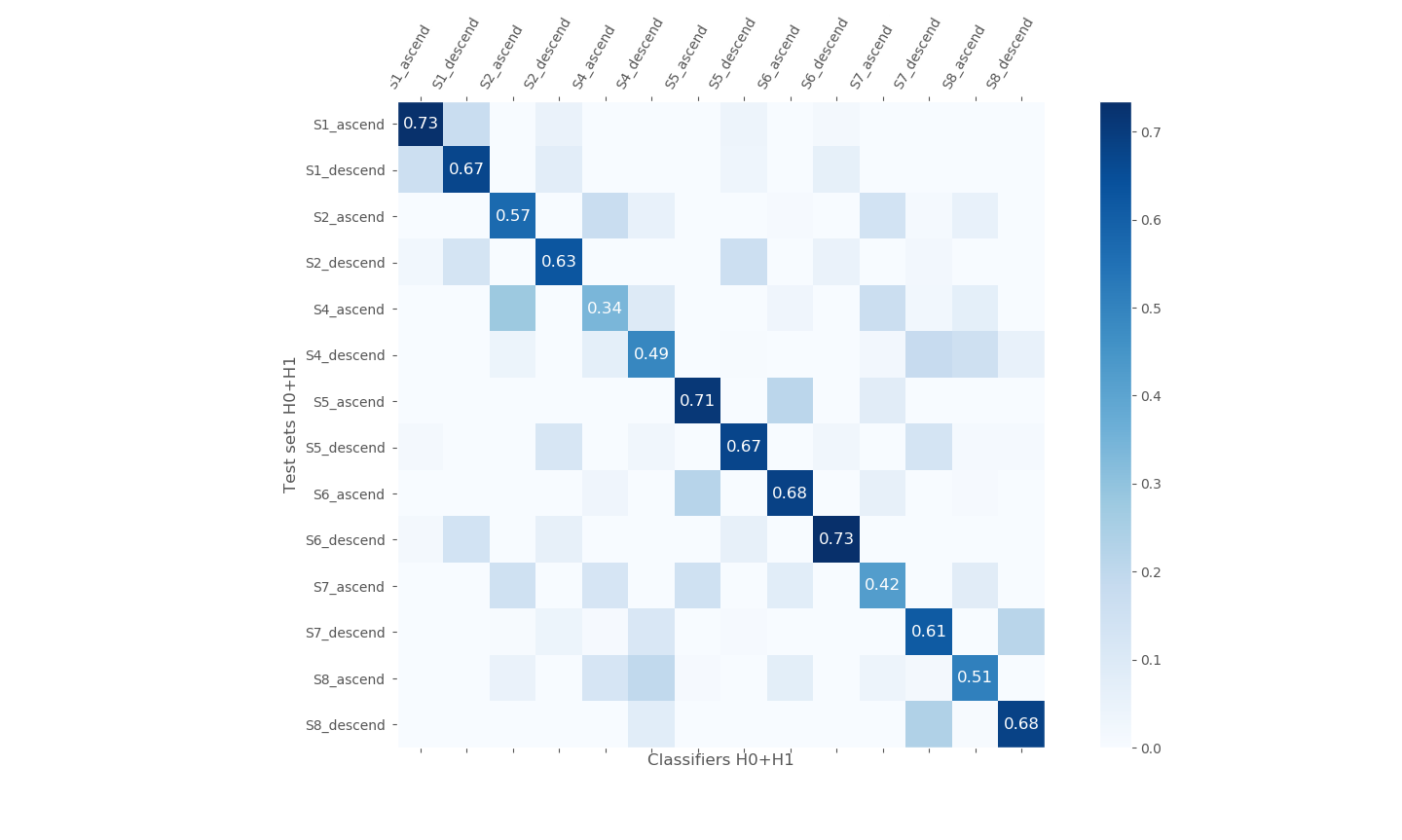}
	\caption{Confusion matrix for the classification of ascending and descending stairs activities with standard contour.}
	\label{fig_activities_confusions_60_acc_standard}
\end{figure}

\begin{figure}
	\centering
	\includegraphics[scale=0.5,clip,trim=0cm 1cm 0cm 0cm]{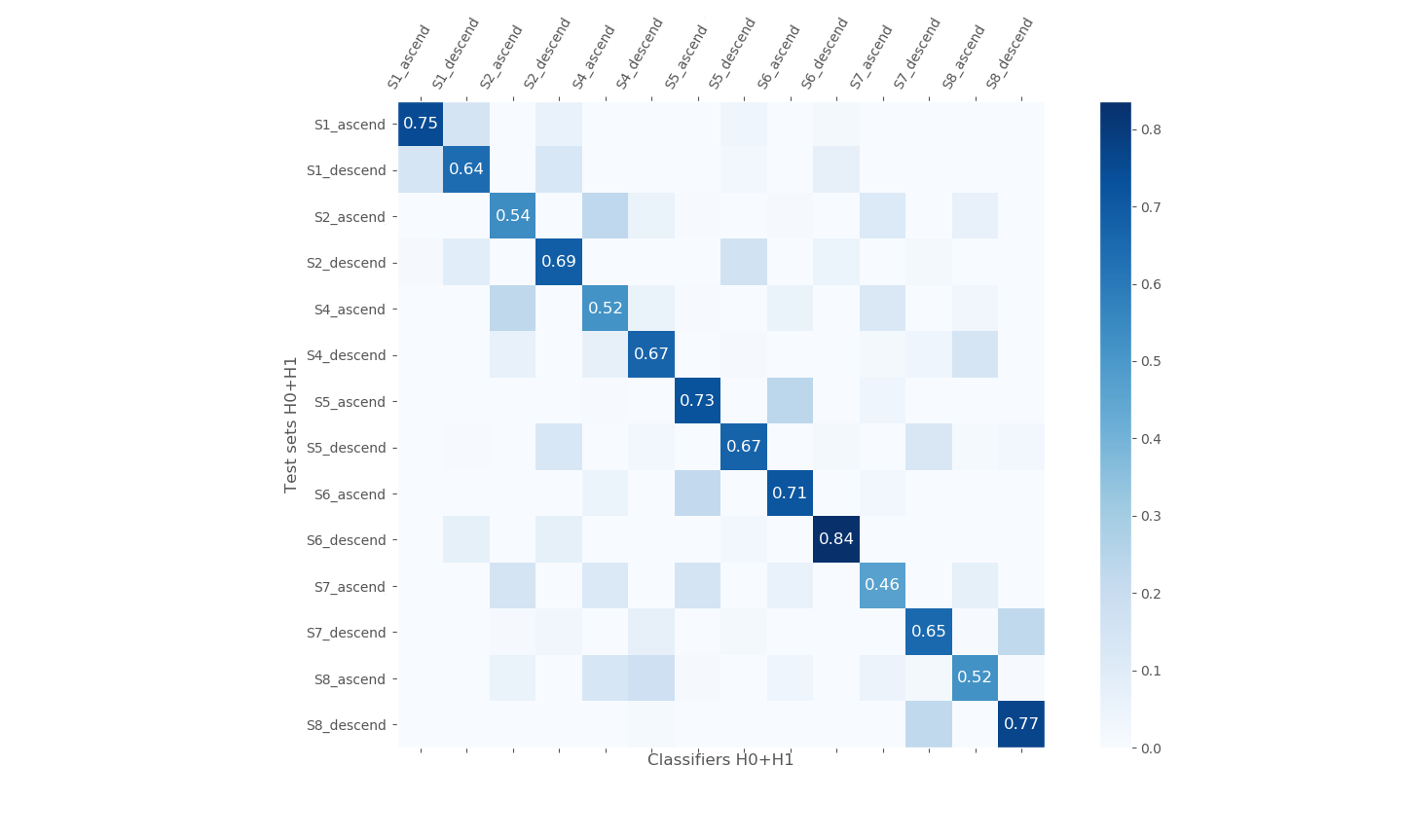}
	\caption{Confusion matrix for the classification of ascending and descending stairs activities with contour coming from density function of Figure \ref{fig_activities_density_contour}.}
	\label{fig_activities_confusions_65_acc_contoured}
\end{figure}

\begin{figure}
	\begin{minipage}{0.99\textwidth}
		\includegraphics[width=0.49\textwidth,clip,trim=3.5cm 1.5cm 1.5cm 2.5cm]{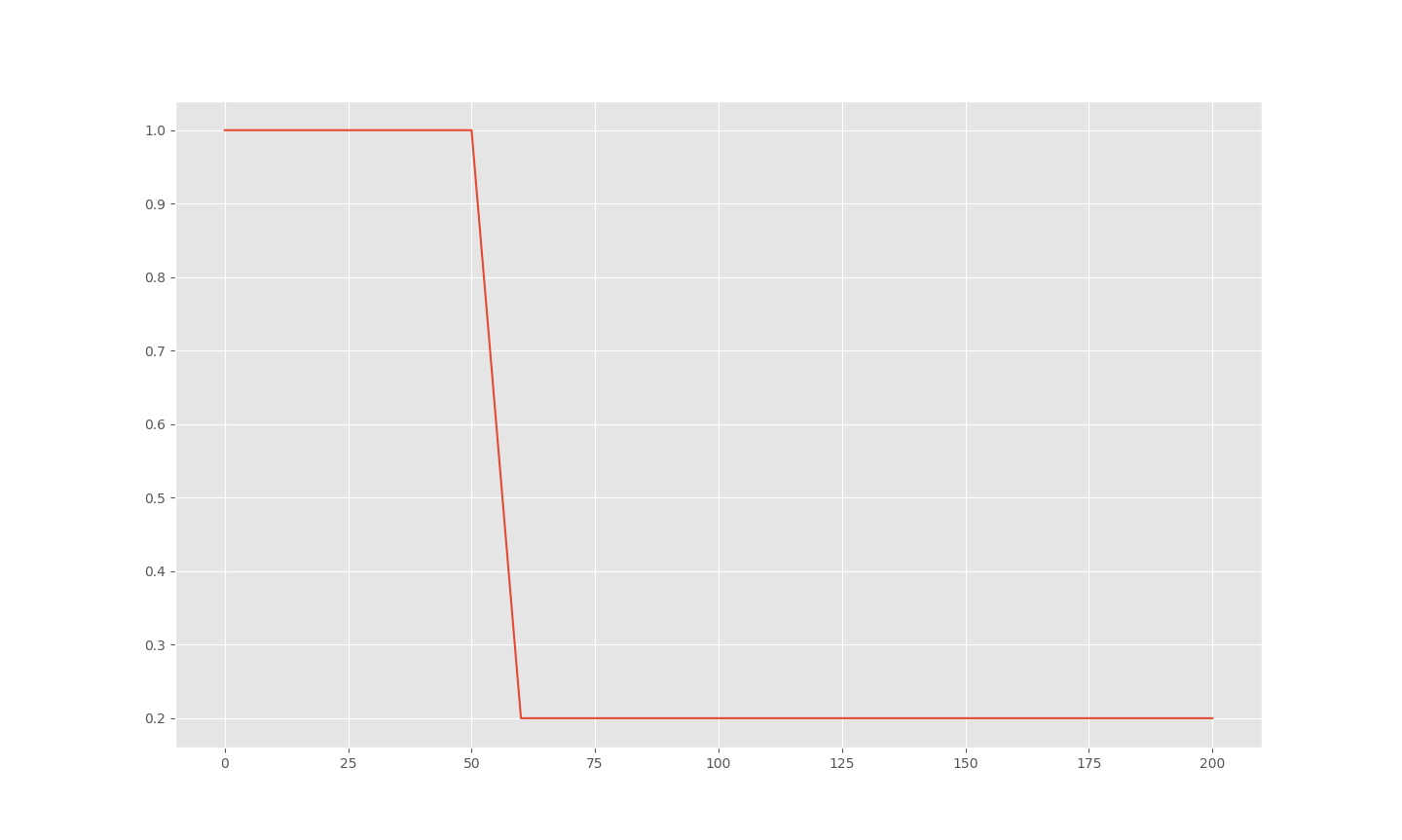}
		\includegraphics[width=0.49\textwidth,clip,trim=3.5cm 1.5cm 1.5cm 2.5cm]{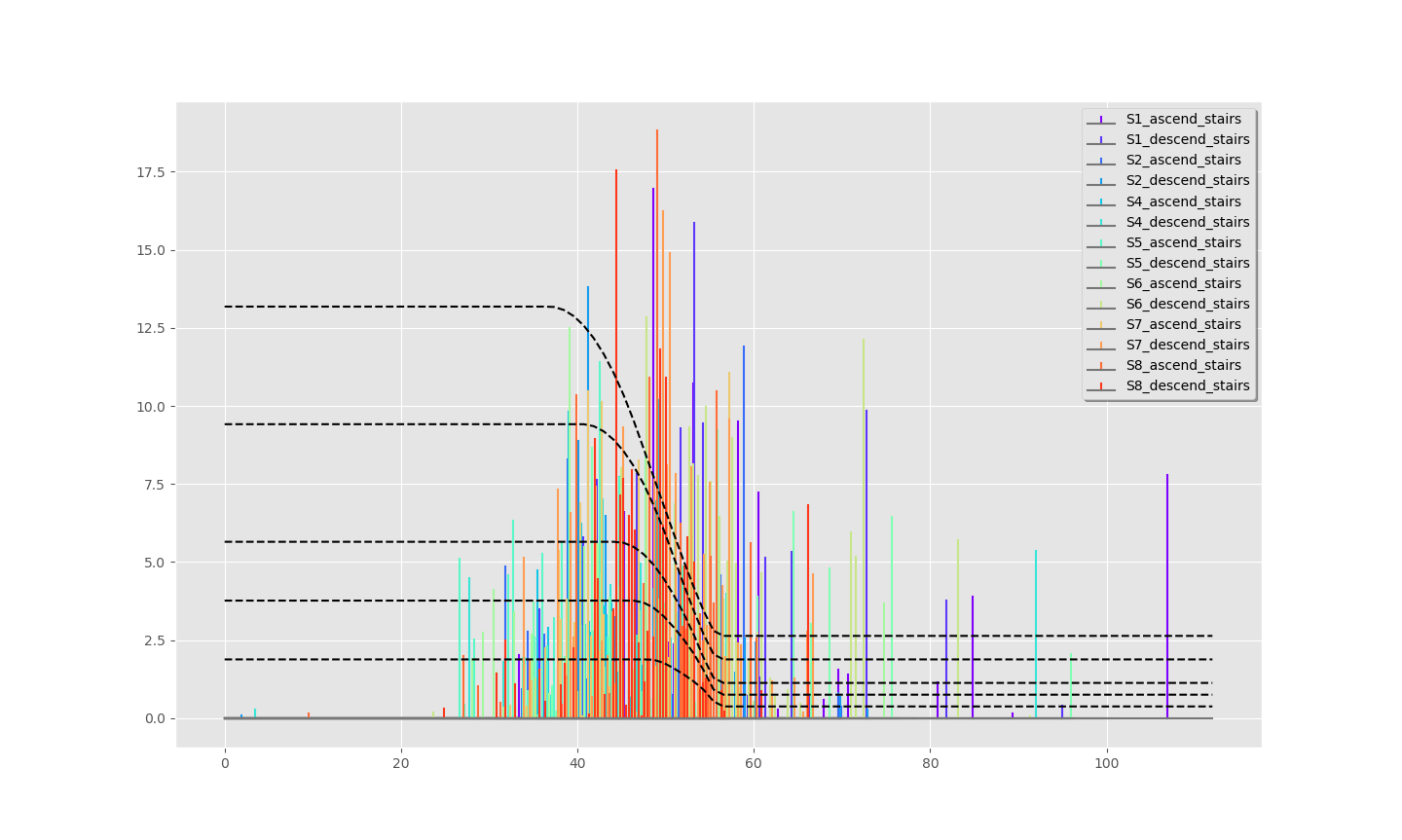}
	\end{minipage}
	\caption{Density function used for $H_1$ signature in the activities classification (left) and contour lines for few values of $\epsilon$ (right). Persistence stems are shown for single data sets.}
	\label{fig_activities_density_contour}
\end{figure}

Result for 20-fold random subsampling cross-validation is shown in Figure \ref{fig_activities_confusions_60_acc_standard} for the standard contour. Overall accuracy is 60\%. We then repeated cross-validation using shift type contour for $H_1$ signature. Contour was obtained from the step function density on left side of Figure \ref{fig_activities_density_contour}. This contour puts more weight on features appearing with larger filtration scales. Contour lines and stem plot are visualized on right side of Figure \ref{fig_activities_density_contour}. Cross-validation results are shown in Figure \ref{fig_activities_confusions_65_acc_contoured}. Overall accuracy increased to 65\%. Note particularly increase in the accuracy of subject 4. Also noteworthy is that ascendings mainly get confused with ascendings of different subjects and the same for descendings. These data thus exhibit clearly different character and using contour makes this difference more pronounced. 

\section{Discussion}
We focused on explaining our data analysis pipeline based on stable rank invariant and persistence contours. Our goal was to demonstrate computationally how one uses this approach. Two following endeavours are the aim of our future research. Statistical stability of stable rank was demonstrated with computational examples. We will continue to develop the statistical theory of this invariant. In particular we will prove central limit theorem and formulate confidence intervals.

It is not surprising that  different  metrics and units are used to measure different things. 
So why in persistence only the bottleneck, interleaving or Gromov-Hausdorff distances are typically used? 
Even for homology based invariants why should the same metric be used for  different homological degrees? It is our strong  belief that metrics should be chosen  depending on the data analysis task at hand and used invariants.
How to choose such  metrics  is 
another direction for further work. We call it {\em contour learning}.
We demonstrated one approach how an appropriate contour can lead to a better classifier. However, selection of the contour was done with trial and error by visually inspecting stem plots and contour lines as shown in Section \ref{contour_usage}. We aim to formulate learning method for selecting contours. Interesting avenue would be linear models. Referring to stem plots and contour lines in Section \ref{contour_usage}, a contour can be interpreted as a weighting of bars in the $1$-dimensional setting, making some bars relatively longer. This should be compared with general linear model $\langle \mathbf{w},\mathbf{x} \rangle+b$, where $\langle \cdot,\cdot \rangle$ is inner product in the appropriate space, $\mathbf{x}$ is data vector, $b$ is bias constant and $\mathbf{w}$ is the weight vector to be determined. Is it possible to discretize our pipeline and learn a contour as a weight vector, leading to an efficient learning algorithm with closed form solution?

\bibliographystyle{plain}
\bibliography{bibliography}
\end{document}